\documentclass[journal,twoside,web]{ieeecolor}
\usepackage{comment}
\usepackage{generic}
\usepackage{cite}
\usepackage{bbm}
\usepackage{amsmath,amssymb,amsfonts}
\usepackage[dvipsnames]{xcolor}
\usepackage{hhline,colortbl,arydshln}
\usepackage{algorithm,algorithmic}
\usepackage{graphicx}
\usepackage{hyperref}
\hypersetup{hidelinks=true}
\usepackage{textcomp}
\usepackage{tabularx,multirow}
\usepackage{diagbox}
\usepackage{soul}
\usepackage{booktabs}
\usepackage{scalerel}
\usepackage{placeins}
\usepackage{xcolor}
\usepackage{makecell}
\usepackage[caption=false,font=footnotesize]{subfig}

\def\BibTeX{{\rm B\kern-.05em{\sc i\kern-.025em b}\kern-.08em
    T\kern-.1667em\lower.7ex\hbox{E}\kern-.125emX}}
    
\newcolumntype{Y}{>{\centering\arraybackslash}X}
\newlength{\charlength}
\begin{document}
\title{{Prism-SQA: An Interpretable and Adaptable Neural Framework for Surface Electromyography Quality Assessment}}

\author{Kuan-Chen Wang, Kai-Chun Liu, Ping-Cheng Yeh, Sheng-Yu Peng, \IEEEmembership{Senior Member, IEEE}, and Yu Tsao, \IEEEmembership{Senior Member, IEEE}
\thanks{This work was supported by the National Science and Technology Council of Taiwan under Grant NSTC 114-2634-F-001-003 and by Academia Sinica under Grants ASGC-111-M01 and AS-IV-115-M01.
 (Corresponding Authors: Kai-Chun Liu; Yu Tsao).}
\thanks{Kuan-Chen Wang is with the Graduate Institute of Communication Engineering, National Taiwan University, Taiwan (email: d12942016@ntu.edu.tw). }
\thanks{Kai-Chun Liu is with the Department of Biomedical Engineering, National Yang Ming Chiao Tung University, Taiwan (email: kcliu33@nycu.edu.tw).}
\thanks{Ping-Cheng Yeh is with the Graduate Institute of Communication Engineering, National Taiwan University, Taiwan (email: pcyeh@ntu.edu.tw).}
\thanks{Sheng-Yu Peng is with the Institute of Electrical and Computer Engineering, National Yang Ming Chiao Tung University, Taiwan (email: speng@nycu.edu.tw).}
\thanks{Yu Tsao is with the Research Center for Information Technology Innovation,
Academia Sinica, Taiwan, and also with the Department of Electrical
Engineering, Chung Yuan Christian University, Taiwan (email: yutsao@as.edu.tw).}
}

\maketitle

\begin{abstract}
Surface electromyography (sEMG) is vulnerable to various contaminants that distort signal morphology and spectral content. Accurate signal quality assessment (SQA) is essential for identifying such degradation and ensuring reliable clinical analyses and decisions. Recent neural network (NN)–based SQA methods achieve accurate quality estimation by learning complex contamination patterns, yet their black-box nature prevents clinicians from understanding or validating the reported quality scores and limits adaptability to application-specific quality definitions without retraining. To address these limitations, we propose Prism-SQA, an interpretable and adaptable neural framework that reformulates SQA as a physiology-aware source-separation and verification process. Prism-SQA decomposes each input signal into a clean sEMG component and five contaminant-specific components using a U-Net with bidirectional long short-term memory. Each separated contaminant component is subsequently examined by a Contaminant Fingerprint Verifier (CFV), which enforces physiological plausibility by comparing the component’s temporal and spectral structure with canonical contaminant signatures. This design allows clinicians to directly inspect how each contaminant affects signal quality, grounding the assessment in transparent, signal-level evidence rather than opaque latent representations. Quality indices computed from the verified components further enable the customization of quality criteria across various clinical contexts without requiring retraining. We evaluate Prism-SQA on continuous quality-score estimation using synthesized noisy sEMG from public Ninapro datasets and on binary quality classification using a clinical dysphagia dataset (SQI dataset). Experimental results demonstrate that Prism-SQA achieves competitive or better performance than contemporary black-box neural methods, providing explicit interpretability and adaptability, and advancing toward practical and clinically aligned sEMG SQA.\footnote{Code: \url{https://github.com/eric-wang135/Prism-SQA}.}
\end{abstract}


\section{Introduction}
\label{sec:introduction}

Surface electromyography (sEMG) is a non-invasive biomedical signal widely used for assessing neuromuscular activity. sEMG supports a wide range of clinical and assistive applications, such as myoelectric prosthesis control \cite{mukhopadhyay2020experimental, cimolato2022emg, shaikh2024toward}, motor-unit analysis \cite{farina2016characterization}, fatigue monitoring \cite{chang2012wireless}, and assessment of movement or neuromuscular disorders \cite{hogrel2005clinical}. However, the quality of sEMG is often compromised by certain types of contaminants, including low-frequency noise (LFN) such as baseline wander (BW) and motion artifacts (MOA), powerline interference (PLI), electrocardiogram (ECG) interference, spike-like artifacts (SPK), and high-frequency noise (HFN). These contaminants can distort waveform morphology and spectral content, leading to inaccurate interpretations and degraded performance in downstream applications \cite{farago2022review, boyer2023reducing}.

Given these vulnerabilities to noise and artifacts, signal quality assessment (SQA) provides an automatic and objective method for estimating quality, which is essential for ensuring reliable sEMG interpretation and downstream analysis \cite{raghu2022automated, farago2022review}. A core component of SQA is quality quantification, which measures the severity of signal degradation and expresses it as a numerical score \cite{farago2022review}. These scores serve as critical decision criteria, determining whether recordings are acceptable for analysis or require denoising and reacquisition \cite{sinderby1995automatic, fraser2011cleanemg}. Quality quantification methods commonly rely on general indices, such as the signal-to-noise ratio (SNR), to measure overall signal integrity \cite{farago2022review}. In addition, contaminant-specific indices are often considered, as the degree of performance degradation depends on the particular types of contaminants present \cite{zhao2024biosignal, sauer2024signal}.

Conventional approaches to computing these quality indices predominantly rely on handcrafted spectral features \cite{sinderby1995automatic, abser2011cleanemg, abser2012cleanemg, chang2020assessment}. These approaches leverage predefined temporal or frequency characteristics to quantify different contaminant types. For example, MOA is typically identified through low-frequency power (below 20 Hz), HFN through power above 800 Hz, and PLI through energy at 50/60 Hz and its harmonics \cite{sinderby1995automatic, abser2012cleanemg, chang2020assessment}. While straightforward to implement and interpret, such fixed frequency-based rules stumble with nonstationary, compound, or low-power contamination, where handcrafted features lose discriminative capability and yield unreliable estimates \cite{abser2011cleanemg, abser2012cleanemg}. This challenge restricts their applicability in complex real-world situations, motivating the development of more robust, data-driven approaches.

Recent advances in deep learning have enabled more powerful sEMG SQA by leveraging the capacity of neural networks (NNs) to model complex patterns of contaminant signals. Oo \textit{et al.} \cite{oo2020signal} and Sauer \textit{et al.} \cite{sauer2024signal} augmented handcrafted features with shallow networks, multi-layer perceptrons (MLP), to estimate quality indices under various contamination scenarios \cite{oo2020signal, sauer2024signal}. Lee \textit{et al.} \cite{lee2024non} proposed QASE-Net, a convolutional–recurrent model that operates directly on raw waveforms and achieves higher estimation accuracy. Despite improved estimation accuracy, two fundamental limitations hinder clinical adoption of NN-based approaches. First, they lack interpretability: black-box NNs produce quality scores without revealing their underlying reasoning, which impedes clinical verification and raises concerns about trustworthiness \cite{salahuddin2022transparency, quinn2022three}. Second, they lack adaptability across applications: SQA for sEMG varies substantially depending on preprocessing pipelines and task-specific signal characteristics \cite{zhao2024biosignal, sauer2024signal}, yet current models are trained to estimate fixed quality indices \cite{lee2024non, sauer2024signal} and cannot readily adapt to new application-specific requirements without retraining or fine-tuning. These limitations underscore the need for more transparent and adaptable SQA frameworks.

To address these limitations, we propose Prism-SQA, an interpretable and adaptable neural framework for sEMG quality assessment. Instead of treating quality estimation as a black-box task, Prism-SQA reformulates it as a physiology-aware decomposition and verification process, ensuring that quality estimation remains aligned with electrophysiological principles. Prism-SQA operates in three tightly connected stages. 
(1) We develop an NN-based separation model that decomposes the input sEMG into an sEMG component and five contaminant-specific components (PLI, LFN, ECG, SPK, and HFN). By isolating individual contaminant contributions, these intermediate outputs offer transparent, signal-level interpretability. The model architecture combines a U-Net \cite{ronneberger2015u} with bidirectional long short-term memory (BiLSTM) \cite{hochreiter1997long} to capture both local waveform features and temporal dependencies. A specialized objective function guides the separation process to optimize performance for SQA, promoting physiologically meaningful decomposition and suppressing implausible contaminant estimates that would otherwise degrade quality interpretation.
(2) A proposed Contaminant Fingerprint Verifier (CFV) enforces physiological plausibility by comparing each separated contaminant component with canonical fingerprints, such as 50/60~Hz peaks for PLI or dominant low-frequency power for LFN \cite{farago2022review,abser2012cleanemg}. Components that fail to meet predefined thresholds are treated as artifacts and discarded, ensuring that SQA relies only on physiologically valid signals.
(3) After CFV processing, quality indices are derived directly from physiologically verified components. This design decouples the computation of quality indices from network training, allowing users to redefine or extend quality indices without requiring retraining or fine-tuning the model. 
As a result, Prism-SQA leverages the powerful capabilities of NNs to achieve accurate SQA while providing interpretable, signal-level assessment that \textcolor{black}{supports flexible adaptation to application-specific requirements.}

We evaluate Prism-SQA through two experiments, which employ synthesized sEMG signals and a clinical database, respectively, to assess its effectiveness, robustness, and adaptability. The first experiment focuses on estimating the quantitative quality score of synthesized noisy sEMG signals. The model is tasked with estimating six quality indices — one overall SNR and five signal-to-contaminant ratios corresponding to PLI, LFN, ECG, SPK, and HFN — across three sub-databases of the Non-Invasive Adaptive Prosthetics (Ninapro) database \cite{atzori2014electromyography}, which include recordings from both intact and transradial-amputee participants. Controlled contamination with known ground-truth levels enables precise evaluation of quality estimation accuracy. Moreover, a zero-shot cross-sub-database evaluation assesses robustness across variations in participant populations and measurement devices. The second experiment performs external validation using the clinical Signal Quality Indices (SQI) dataset for dysphagia assessment \cite{cuadros2022automatic}, focusing on the task of quality classification under clinical conditions. 
This experiment evaluates Prism-SQA’s ability to generalize beyond the controlled synthetic setting, \textcolor{black}{while the exploratory adaptation analysis illustrates how the derived quality indices can be flexibly modified to reflect alternative dataset-specific quality definitions.} Together, these experiments demonstrate that Prism-SQA achieves strong and consistent performance across diverse datasets, populations, and tasks, while providing physiologically grounded interpretability—marking an important step toward trustworthy sEMG SQA in clinical and rehabilitation applications.

The main contributions of this study are as follows.
\begin{itemize}

\item \textbf{A physiology-aware and interpretable framework for sEMG SQA:} We introduce a new paradigm for sEMG SQA that transforms black-box estimation into a physiology-aware source-separation and verification pipeline. This approach yields physiologically interpretable signal components, enabling direct inspection and validation of contamination patterns in sEMG.

\item \textbf{Enhanced adaptability across applications through decoupled quality index computation}: By decoupling quality index computation from model training, Prism-SQA allows flexible, task-specific customization of quality indices without network retraining. This design supports diverse experimental and clinical requirements, enabling the use of suitable statistical measures or signal components to quantify quality that represents the extent of contamination affecting preprocessing steps or application performance.

\item \textbf{Comprehensive evaluation across multiple datasets}: 

Extensive experiments on three Ninapro subsets and one clinical SQI dataset demonstrate the robustness of Prism-SQA across varied signal conditions, subject populations, and application scenarios.

\end{itemize}

\textcolor{black}{Key abbreviations and notations used throughout this paper are summarized in Table~\ref{tab:abbr} to facilitate reading.}

\section{Materials}
\label{sec:materials}
\subsection{Datasets and preprocessing}

\subsubsection{sEMG database}

This study employs sEMG data from multiple open-access datasets, including three subsets from the Ninapro database \cite{atzori2014electromyography} and the SQI dataset \cite{cuadros2022automatic} for dysphagia assessment. The Ninapro database is used to generate noisy sEMG data with corresponding ground-truth quality indices for the quality score estimation experiment, whereas the SQI dataset is used for the clinical quality classification experiment.

\paragraph{Ninapro database}
The Ninapro database \cite{atzori2014electromyography} provides high-quality sEMG recordings and serves as the clean data source for generating paired clean and noisy signals. We use three subsets—DB2, DB3, and DB4—containing upper-limb sEMG signals collected from 40 intact subjects (DB2), 11 transradial amputees (DB3), and 10 intact subjects (DB4). DB2 and DB3 were recorded using Delsys Trigno electrodes, while DB4 employed Cometa electrodes.
All three subsets include three exercise sessions. DB2 and DB3 share Exercises B, C, and D, whereas DB4 contains Exercises A, B, and C. We adopt Exercise B from DB2 and Exercise C from DB2, DB3, and DB4. During these sessions, subjects performed 17 and 22 types of movements, respectively. Each movement was repeated six times, with each repetition lasting 5 s, followed by a 3-s rest period. Previous studies have demonstrated that sEMG data in Ninapro exhibit good quality \cite{chang2020assessment} and can serve as clean sEMG data after appropriate filtering \cite{machado2021deep, wang2023ecg, zhang2023semg}.

\paragraph{SQI dataset}
\label{para:SQI dataset}
For the quality classification task, we use the SQI dataset \cite{cuadros2022automatic}, an open-access sEMG SQA dataset collected from patients with dysphagia during swallowing tests. The signals were labeled by experts as either good or poor quality. The dataset consists of three subsets, referred to here as SQI-DB1, SQI-DB2, and SQI-DB3. After excluding the 8 purely silent poor-quality segments from SQI-DB1, the final evaluation sets contained 152/144, 50/50, and 50/50 good-/poor-quality segments from SQI-DB1, SQI-DB2, and SQI-DB3, respectively. \textcolor{black}{These segments originated from 39, 20, and 20 subjects.} We exclude silent segments as they can be trivially identified as poor-quality signals by checking whether the signal power is zero. Thus, including them would not provide meaningful insight into the performance of SQA approaches. The remaining poor-quality signals primarily contain realistic contaminants such as PLI, ECG, and MOA.

\begin{table}[t!]
\centering
\scriptsize
\caption{\textcolor{black}{Summary of key abbreviations and notations used in this paper.}}
\label{tab:abbr}
\begin{tabular}{p{0.23\columnwidth}p{0.67\columnwidth}}
\toprule
\textbf{Abbreviation} & \textbf{Full name} \\
\midrule

SQA & Signal quality assessment \\
CFV & Contaminant fingerprint verifier \\
NN & Neural network \\

\midrule
sEMG & Surface electromyography \\
PLI & Powerline interference \\
LFN & Low-frequency noise \\
ECG & Electrocardiogram \\
SPK & Spike-like artifact \\
HFN & High-frequency noise \\
MOA & Motion artifact \\

\midrule
SNR & Signal-to-noise ratio \\
SPR & Signal-to-PLI ratio \\
SLR & Signal-to-LFN ratio \\
SER & Signal-to-ECG ratio \\
SSR & Signal-to-SPK ratio \\
SHR & Signal-to-HFN ratio \\

\midrule
Ninapro & Non-invasive adaptive prosthetics dataset \\
SQI & Signal quality indices \\
DB2/DB3/DB4 & Ninapro subsets \\
SQI-DB1/2/3 & SQI dataset subsets \\

\midrule
SD-SDR & Scale-dependent signal-to-distortion ratio \\
$\mathcal{L}_{1}/\mathcal{L}_{\text{SD-SDR}}/\mathcal{L}_{\text{cls}}$
& L1, SD-SDR, and component-presence classification losses \\

\bottomrule
\end{tabular}
\end{table}

\begin{table}[th!]
\centering
\scriptsize
\caption{\textcolor{black}{Data sources and conditions used for training/validation and quality-score estimation testing.}}
\label{tab:mismatch}
\setlength{\tabcolsep}{2.5pt}
\renewcommand{\arraystretch}{1.05}
\begin{tabular}{
    p{0.09\columnwidth}
    p{0.4\columnwidth}
    p{0.2\columnwidth}
    p{0.23\columnwidth}}
\toprule
\textbf{Signal} & \textbf{Source} & \textbf{Train/Val} & \textbf{Test} \\
\midrule

sEMG &
Ninapro~\cite{atzori2014electromyography} &
\makecell[l]{DB2; \\25/5 subjects;\\Ex.~B; Ch.~1--8} &
\makecell[l]{DB2/DB3/DB4;\\ 10/11/10 subjects;\\Ex.~C; Ch.~11} \\

\midrule

\multirow{3}{*}{LFN}
& BW, EM: MIT-BIH NSTDB~\cite{moody1984noise}
& Ch.~1 & Ch.~2 \\

& MOA: Machado \textit{et al.}~\cite{machado2021deep}
& Ch.~1--8 & Ch.~9--12 \\

& Step function, low-pass WGN
& Simulated & Simulated \\

\midrule

PLI &
\makecell[l]{50/60-Hz sinusoids \\ with harmonics} &
Simulated & Simulated \\

\midrule

ECG &
MIT-BIH NSRD~\cite{goldberger2000physiobank} &
14 subjects &
\makecell[l]{4 subjects \\ (19090, 19093, \\ 19140, 19830)} \\

\midrule

SPK &
Delta functions, MUAPs &
Simulated & Simulated \\

\midrule

HFN &
WGN, band-passed WGN &
Simulated & Simulated \\

\bottomrule
\end{tabular}
\end{table}

\paragraph{Preprocessing}

Clean sEMG data from the Ninapro database were preprocessed using a fourth-order Butterworth bandpass filter with cutoff frequencies of 20 Hz and 500 Hz. The filtered signals were then downsampled from 2 kHz to 1 kHz, segmented into 2-second windows, and silent segments were removed. Although the bandpass filter suppresses components below 20 Hz and above 500 Hz, this range preserves the most relevant physiological information and is widely adopted in sEMG research and applications~\cite{xiaojing2011feature, thongpanja2013mean, cao2022control}.

For the SQI dataset, signals from SQI-DB1, SQI-DB2, and SQI-DB3 were downsampled from 2 kHz, 2 kHz, and 5 kHz to a 1 kHz sampling rate, respectively. This resampling ensures compatibility with the NN models trained on Ninapro data, maintaining consistent temporal resolution across datasets.

\subsubsection{sEMG contaminants dataset}

Five contaminant types were considered for generating noisy sEMG. To ensure consistency, all contaminant signals were resampled to 1 kHz to match the sampling rate of the sEMG data. The characteristics and sources of these contaminants are described below.

\paragraph{LFN} 
We employed six types of LFN: BW, electrode motion (EM), MOA, a step function, and two variants of low-pass-filtered white Gaussian noise (WGN). BW and EM were sourced from the MIT-BIH Noise Stress Test Database (NSTDB) \cite{moody1984noise}, which has been widely used as a representative source of LFN contaminants in sEMG studies \cite{yadav2023noise, wang2024trustemg}. Additionally, MOA data were obtained from \cite{machado2021deep}, which were generated by tapping the electrodes ten times at 1-s intervals without muscle contraction to induce slight electrode displacement. These signals were further smoothed using a 51-sample moving average filter \cite{machado2021deep}. The remaining three LFN types were synthetically generated: a step-function signal with a duration of 0.5 s, and two low-pass-filtered WGN signals obtained using Butterworth filters with cutoff frequencies of 20 Hz and 40 Hz, respectively \cite{sauer2024signal}.

\paragraph{PLI} 
PLI was simulated using a sinusoidal waveform at either 50 Hz or 60 Hz~\cite{machado2021deep,boyer2023reducing}. For better realism, we randomly incorporated a frequency drift of up to 1.5 Hz~\cite{mateo2008neural} and generated the first and second harmonics of the PLI \cite{keshtkaran2014fast}.

\paragraph{ECG} 
ECG data were obtained from the MIT-BIH Normal Sinus Rhythm Database (NSRD), which contains 24 hours of ECG recordings from 14 subjects~\cite{goldberger2000physiobank}. This dataset is commonly used in sEMG denoising and contaminant-type identification research as a realistic source of ECG interference~\cite{machado2021deep,wang2023ecg,wang2024trustemg}. To discard potential noise within the ECG data, we applied a preprocessing pipeline consisting of a 1 Hz high-pass filter to remove BW, a 60 Hz notch filter for PLI, and a 200 Hz low-pass filter to suppress HFN~\cite{mccool2014identification,abdelazez2018detection}.

\paragraph{SPK} 
We considered the SPK of two morphologies. The first type was unspecific instantaneous spikes, generated as delta pulses with a 5 ms duration~\cite{sauer2024signal}. The second type was physiologically motivated and modeled individual motor unit action potentials (MUAPs) that occur regularly and dominate the remaining sEMG content~\cite{sauer2024signal}. Each MUAP was simulated using a second-order Gaussian function~\cite{li2017wavelet}.

\paragraph{HFN} 
HFN was modeled as both WGN and bandpass-filtered WGN. The latter was obtained by applying a fourth-order Butterworth bandpass filter with cutoff frequencies of 100 Hz and 400 Hz~\cite{sauer2024signal}.

\paragraph{Combined contaminant}
We generated compound contaminants by combining multiple contaminant types to introduce more challenging SQA conditions~\cite{ma2020emg, sauer2024signal}. These mixtures contained two to five noise components, with each component contributing an equal amount of signal power.

\subsubsection{Noisy sEMG dataset}
The noisy sEMG dataset used in our experiments was synthesized by scaling and linearly adding contaminants to clean sEMG signals at specified SNRs, calculated as:

\begin{equation}
\label{eq:SNR}
\small
\text{SNR (dB)} = 10\log_{10}\left(\frac{P_{\text{signal}}}{P_{\text{noise}}}\right) = 10\log_{10}\frac{\sum_{n=1}^{N}x_{s}[n]^2}{\sum_{n=1}^{N}x_{n}[n]^2},
\end{equation}
where \(P_{\text{signal}}\) and \(P_{\text{noise}}\) denote the power of the clean sEMG signal and the contaminant, respectively, and \(x_s\) and \(x_n\) denote their waveforms with length \(N\). In this study, we set the maximum value of SNR to 30 dB, which is often used to indicate a clean signal in previous studies \cite{chang2020assessment,sauer2024signal}.   

\begin{figure*}[tbh!]
    \centering
    \includegraphics[width=.9\textwidth]{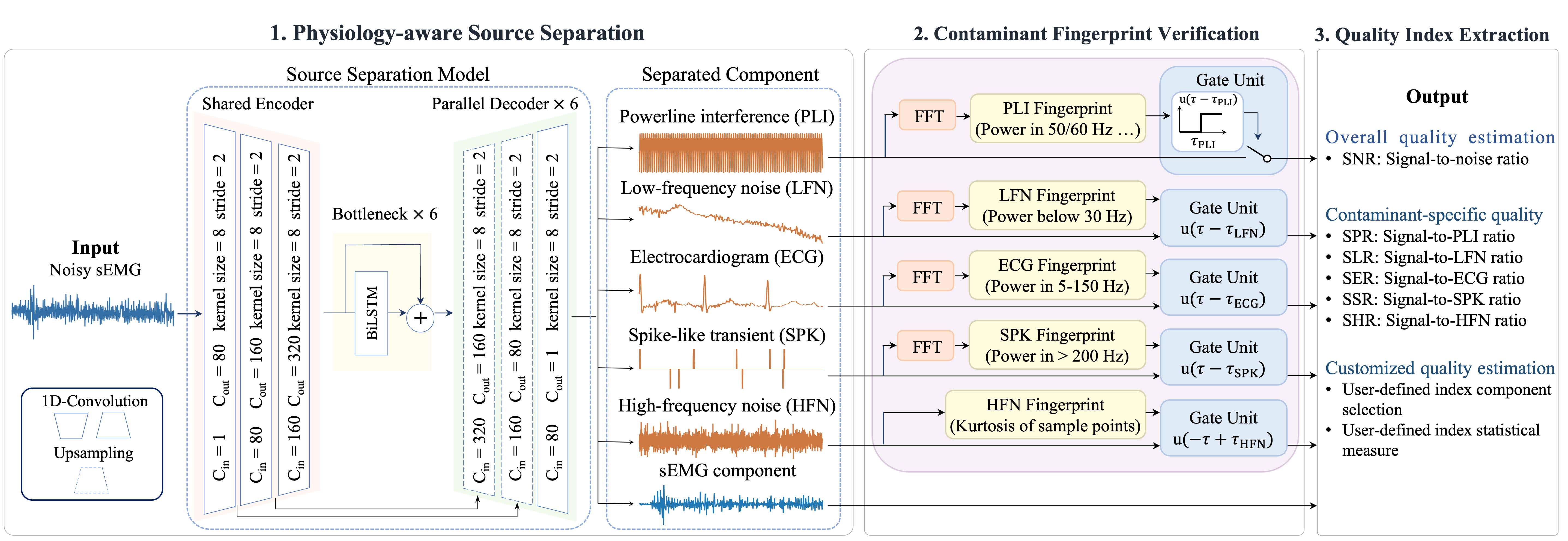}
    \caption{The framework of the proposed Prism-SQA, consisting of three stages: physiology-aware source separation, contaminant fingerprint verification, and quality index extraction.}
    \label{fig: Model structure}
\end{figure*}

\subsection{Data preparation}

\paragraph{Training and validation sets}

To train the source separation model, we prepared noisy sEMG datasets using data from Ninapro DB2, which provides the largest subject pool. Clean sEMG was extracted from Channels 1–8 during Exercise B from 25 subjects for training and 5 subjects for validation. This resulted in 38787 and 10221 clean sEMG segments for the training and validation sets, respectively. 

A dynamic data mixing strategy was employed to enhance training diversity. During training, noisy sEMG segments were generated on the fly for each clean segment by randomly sampling the number and types of contaminants, as well as the SNR. The number of contaminants \(K\) was drawn uniformly from 0 to 5, and \(K\) contaminant types were randomly selected from PLI, LFN, ECG, SPK, and HFN. The SNR was sampled uniformly from –20 to 30 dB. The selected contaminants were scaled and added to the clean sEMG signal to ensure that the resulting noisy signal achieved the sampled SNR.

For each noisy sEMG segment \(y \in \mathbb{R}^{1 \times N}\), the corresponding clean signal and contaminant waveforms (with zero vectors assigned to absent contaminants) were concatenated as the source-separation ground truth, resulting in a matrix \(G \in \mathbb{R}^{6 \times N}\). To maintain consistent scaling across datasets, each noisy input segment was normalized to the range of [–1, 1] by dividing by its maximum absolute amplitude, and the corresponding ground-truth waveforms were scaled by the same factor to preserve proportional relationships. To evaluate both contaminant-specific and overall quality, five contaminant-specific indices—signal-to-PLI ratio (SPR), signal-to-LFN ratio (SLR), signal-to-ECG ratio (SER), signal-to-SPK ratio (SSR), and signal-to-HFN ratio (SHR)—were computed as SNRs between the clean signal and each contaminant. These five indices, together with the overall SNR, formed the ground-truth quality targets \(g \in \mathbb{R}^{6 \times 1}\).

\paragraph{Test sets for quality score estimation}

We rigorously assessed the proposed approach by constructing test sets using datasets, recording setups, and contamination conditions that differ from those used during training. Table~\ref{tab:mismatch} summarizes the data sources, acquisition configurations, and contaminant types used in the training/validation and testing phases.

Clean sEMG signals for the test sets were obtained from Channel 11 during Exercise C, recorded from 10, 11, and 10 subjects in Ninapro DB2, DB3, and DB4 datasets, respectively. This yielded 3180, 3177, and 3669 clean segments. These test sets introduce different levels of mismatch relative to the training sEMG data: 
all three differ in subjects, exercises, and channels; 
DB3 additionally differs in subject condition (transradial amputees), and DB4 differs in recording devices (Cometa electrodes). 

Each clean segment was then contaminated under five SNR levels (–20, –10, 0, 10, and 20 dB) and with one to five contaminant types, with all five contaminant-number conditions represented equally. Moreover, each contaminant type (PLI, LFN, ECG, SPK, and HFN) was included with equal frequency in the synthesized test data. Additionally, we included clean sEMG segments alongside synthesized noisy data in the test sets to assess whether the model could accurately estimate quality for clean signals. In total, the test sets comprised 82680 (3180 × 5 SNRs × 5 contaminant-number conditions + 3180 clean), 82602 (3177 × 5 × 5 + 3177 clean), and 95394 (3669 × 5 × 5 + 3669 clean) segments for DB2, DB3, and DB4, respectively. The corresponding ground-truth waveforms and quality indices were generated in the same way as for the training set.

\paragraph{Test sets for quality classification experiment}

For the quality classification task, we directly used the SQI dataset, a clinical dataset containing both high- and low-quality sEMG signals, to evaluate the proposed framework under practical conditions.

\section{Methodology}
\label{sec:methodology}
\subsection{Proposed method}
The pipeline of the proposed Prism-SQA, shown in Fig.~\ref{fig: Model structure}, consists of three stages:

\subsubsection{Physiology-aware source separation}
\label{NN details}
The first stage aims to decompose the input sEMG signal into an sEMG component and five contaminant-specific components (LFN, PLI, ECG, SPK, and HFN), providing physiologically meaningful intermediate outputs for downstream SQA. To achieve this, we develop a source separation model tailored for downstream quality estimation. We introduce its design in two parts: model architecture and its objective function.

\paragraph{Model architecture}
The proposed source separation model is based on an autoencoder design, utilizing a U-Net \cite{ronneberger2015u} backbone with BiLSTM layers \cite{hochreiter1997long} in the bottleneck. The model takes a noisy sEMG waveform as input and produces six output components corresponding to the clean sEMG signal and five contaminant-specific signals. This architecture enables effective decomposition of contaminant sources with diverse temporal and spectral characteristics \cite{takahashi2018mmdenselstm, defossez2019music}.

\textbf{Encoder.} The encoder comprises three one-dimensional convolutional layers with a kernel size of 8 and a stride of 2, progressively reducing temporal resolution while expanding feature dimensionality. Specifically, the encoder transforms the input waveform of size \(d \times 1\) into feature representations with dimensions \(\frac{d}{2} \times 80\), \(\frac{d}{4} \times 160\), and \(\frac{d}{8} \times 320\). These layers mainly capture local features of the input signal, such as waveform peaks, zero crossings, and power variations, which are informative for identifying and separating waveform components \cite{huang1998empirical}. Each convolutional layer is followed by 1D batch normalization and a ReLU activation.

\textbf{BiLSTM bottleneck.} To incorporate long-range temporal dependencies, the encoder output is passed through a stack of six BiLSTM layers, each with a hidden size of 320 (matching the encoder feature dimension). Residual connections between layers facilitate the flow of gradients and improve training stability. By modeling temporal context across the entire signal, the BiLSTM bottleneck complements the local feature extraction of the convolutional layers. This design enhances the model's ability to separate contaminants with extended temporal structures, such as ECG and PLI~\cite{alizadegan2025comparative}.

\textbf{Decoder.} The decoder mirrors the encoder, consisting of two upsampling modules followed by a transposed convolutional layer. Each upsampling module consists of a transposed convolution layer (stride of 2) and a convolution layer (stride of 1), both of which are followed by 1D batch normalization and ReLU activations. Skip connections bridge encoder and decoder layers at corresponding resolutions, preserving fine-grained details that might otherwise be lost during downsampling \cite{ronneberger2015u}.
Finally, six parallel decoder heads reconstruct the separated components, with each head implemented as a transposed convolution with a single kernel. The six outputs are then concatenated along the channel dimension to form a tensor of size $6 \times d$, corresponding to the sEMG and five contaminant waveforms.

\paragraph{Objective function}

To enable interpretable SQA, the separation model is trained with an objective function specifically designed to promote physiologically meaningful decomposition. Unlike standard source separation objectives that focus solely on reconstruction accuracy, our loss function explicitly addresses two critical requirements: (1) maintaining correct amplitude relationships between sEMG and contaminants, since contamination magnitude indicates signal quality \cite{farago2022review}, and (2) reliably detecting contaminant presence to avoid false outputs that could distort quality estimates. We implement these requirements by combining the standard L1 loss with a scale-dependent signal-to-distortion ratio (SD-SDR) loss \cite{le2019sdr} and a component-presence classification loss.

The overall objective is formulated as:
\begin{equation}
\small
\mathcal{L}
= \lambda_{1}\, \mathcal{L}_{1}
+ \lambda_{2}\, \mathcal{L}_{\text{SD\!-\!SDR}}
+ \lambda_{3}\, \mathcal{L}_{\text{cls}},
\label{eq:total_loss}
\end{equation}
where \(\lambda_1,\lambda_2,\lambda_3>0\) weight the reconstruction loss $\mathcal{L}_{1}$, the scale-dependent distortion loss $\mathcal{L}_{\text{SD\!-\!SDR}}$, and the component-presence classification loss $\mathcal{L}_{\text{cls}}$, respectively.

Let \(\mathbf{X} \in \mathbb{R}^{C \times N}\) and \(\hat{\mathbf{X}} \in \mathbb{R}^{C \times N}\) 
denote the ground-truth and estimated component matrices, where 
\(C\) is the total number of signal components (one sEMG component and \(C-1\) contaminant components) 
and \(N\) is the number of time samples. 
The \(c\)-th row of \(\mathbf{X}\), denoted \(\mathbf{x}_c \in \mathbb{R}^{N}\), 
and the corresponding row of \(\hat{\mathbf{X}}\), denoted \(\hat{\mathbf{x}}_c\), 
represent the ground-truth and estimated waveforms of the \(c\)-th component. 
The first row corresponds to the sEMG component, and the remaining rows correspond to contaminant components. 

\textbf{L1 loss.} The $\mathcal{L}_{1}$ loss encourages the output components to approximate the ground truth by minimizing the pointwise difference between the estimated and ground-truth components:
\begin{equation}
\small
\mathcal{L}_{1} = \frac{1}{CN} \sum_{c=1}^{C} \sum_{t=1}^{N} \bigl| \mathbf{x}_{c}[t] - \hat{\mathbf{x}}_{c}[t] \bigr|.
\end{equation}

\textbf{SD-SDR loss \cite{le2019sdr}.} We employ the
$\mathcal{L}_{\text{SD-SDR}}$ to ensure that the separated components are not only structurally accurate but also preserve the correct amplitude relationships, a property crucial for reliable quality estimation. \textcolor{black}{SD-SDR retains the
reconstruction error relative to the original target in the denominator, thereby penalizing amplitude scaling errors.} The loss is computed over channels with existing components whose ground-truth waveform is nonzero, and is defined as:
\begin{equation}
\small
\mathcal{L}_{\text{SD-SDR}}
= -\,\frac{1}{|\mathcal{C}_{\text{exist}}|} 
\sum_{c \in \mathcal{C}_{\text{exist}}} 
10 \log_{10} 
\frac{\| \alpha_c\, \mathbf{x}_c \|_2^2}
{\| \mathbf{x}_c - \hat{\mathbf{x}}_c \|_2^2},
\end{equation}
where \(\mathcal{C}_{\text{exist}} = \{\, c \;|\; \mathbf{x}_c \neq \mathbf{0} \,\}\) is the set of channels of existing components, \(|\mathcal{C}_{\text{exist}}|\) denotes the number of existing components, and \(\alpha_c = \frac{\langle \mathbf{x}_c,\, \hat{\mathbf{x}}_c \rangle}{\|\mathbf{x}_c\|_2^2}\) is a scaling factor that projects the estimate \(\hat{\mathbf{x}}_c\) onto the target \(\mathbf{x}_c\).

\textbf{Component-presence classification loss.} The $\mathcal{L}_{\text{cls}}$ aims to address spurious artifacts from absent contaminant components that can be misinterpreted as real contaminants, leading to erroneously low quality scores. The component-presence classification loss encourages near-zero outputs for absent components and strong activations only when contaminants are present. 
For each contaminant channel \(c\) (excluding the sEMG channel), we define a confidence score \(\hat{z}_c\) based on the total magnitude of the estimated waveform:
\begin{equation}
\small
\hat{z}_c = 2\sigma\!\left( \sum_{t=1}^{N} \bigl| \hat{\mathbf{x}}_{c}[t] \bigr| \right) - 1,
\end{equation}
where \(\sigma(\cdot)\) is the sigmoid function. Because the input to the sigmoid is non-negative, \(\sigma(\cdot)\) outputs values in \((0.5,\,1)\).  
Applying the transformation \(2\sigma(\cdot)-1\) maps this range to \((0,\,1)\), where values near \(0\) indicate negligible estimated energy (contaminant absent) and values near \(1\) indicate substantial energy (contaminant present).

The corresponding binary ground-truth label is defined as:
\begin{equation}
\small
z_c = \mathbbm{1}\left( \sum_{t=1}^{N} |\mathbf{x}_{c}[t]| > 0 \right),
\end{equation}
where $\mathbbm{1}(\cdot)$ is the indicator function, returning 1 when the contaminant component contains any non-zero energy and 0 otherwise. The classification loss is formulated as the binary cross-entropy between estimated and ground-truth labels for all five contaminant channels:
\begin{equation}
\small
\mathcal{L}_{\mathrm{cls}}
= -\,\frac{1}{C-1} \sum_{c=2}^{C}
\Bigl[\, z_c \log(\hat{z}_c) + (1 - z_c)\log\!\bigl(1 - \hat{z}_c \bigr) \Bigr],
\end{equation}
where \(C-1\) denotes the number of contaminant channels.

\begin{table}[t]
\centering
\scriptsize
\caption{\textcolor{black}{Fingerprints and thresholds used in the CFV.}}
\label{tab:cfv}

\renewcommand{\arraystretch}{1.5}
\begin{tabular}{cll}
\toprule
\textbf{Contaminant} & \textbf{Fingerprint} & \textbf{Verification Criterion} \\ 
\midrule

PLI &
\makecell[l]{Narrowband concentration\\(50/60~Hz and harmonics)} &
$P_{48:52,\,58:62\text{Hz...}} > \tau_{\text{PLI}}$ \\

LFN &
Low-frequency dominance &
$P_{0:30\text{Hz}} > \tau_{\text{LFN}}$ \\

ECG &
Mid-band concentration &
$P_{5:150\text{Hz}} > \tau_{\text{ECG}}$ \\

SPK &
High-frequency impulsivity &
$P_{200:500\text{Hz}} > \tau_{\text{SPK}}$ \\

HFN &
Kurtosis of waveform &
$|\kappa| < \tau_{\text{HFN}}$ \\

\bottomrule
\end{tabular}

\vspace{.5mm}
\parbox{0.88\columnwidth}{\scriptsize
{\tiny $P_{f_1:f_2}$} denotes the power ratio within the frequency band
{\tiny $[f_1,f_2]$}, $\kappa$ denotes waveform kurtosis, and
$\tau$ denotes the verification threshold.
}
\end{table}
\renewcommand{\arraystretch}{1.0}

\subsubsection{Contaminant fingerprint verification}

Based on the interpretable outputs of the separation model, we introduce CFV, a knowledge-based verification module designed to ensure that the separated components conform to electrophysiological characteristics before quality estimation. CFV evaluates each separated contaminant according to its fingerprint—a distinct spectral or temporal pattern derived from established sEMG SQA principles. Fingerprints and verification criteria for each contaminant type are summarized in Table~\ref{tab:cfv}. 

\textcolor{black}{CFV extracts two types of handcrafted features as fingerprints. For PLI, LFN, ECG, and SPK, we use the power ratio \(P_{f_1\!:\!f_2}\) within a specific frequency band \([f_1,f_2]\), computed from the Fast Fourier Transform (FFT) spectrum \cite{farago2022review}:}

\begin{equation}
\small
\textcolor{black}{P_{f_1\!:\!f_2}
= \sum_{f=f_1}^{f_2}|X(f)|^2
\big/
\sum_f |X(f)|^2,}
\end{equation}
\textcolor{black}{where \(X(f)\) is the Fourier transform of the separated component. The frequency bands are set to 0--30~Hz for LFN, 50/60~Hz and harmonics (\(\pm2\)~Hz) for PLI, 5--150~Hz for ECG, and 200--500~Hz for SPK.} \textcolor{black}{For HFN, we apply waveform kurtosis \(\kappa\), where larger absolute values indicate non-Gaussian characteristics.}

Each fingerprint is compared with a threshold \(\tau\) in a gating unit to verify whether the component exhibits the expected spectral or temporal structure. Components that satisfy their thresholds are retained, while those that fail are treated as artifacts and replaced with zero vectors. In this study, thresholds \(\tau_{\text{PLI}}, \tau_{\text{LFN}}, \tau_{\text{ECG}}, \tau_{\text{SPK}}\), and \(\tau_{\text{HFN}}\) were set to 0.3, 0.3, 0.3, 0.3, and 0.6, respectively. These values were chosen based on the scale and interpretation of each contaminant fingerprint measure, rather than by optimizing performance on the test set. The values were selected to avoid two degenerate cases: allowing nearly all separated components to pass verification, or rejecting most reconstructed contaminant components. To further justify this choice, a sensitivity analysis on Ninapro DB2 is presented in Section \ref{subsec:sensitivity of CFV}, showing that the performance remains stable across a broad range of threshold values and only degrades under extreme settings.

CFV is activated only when the preliminary SNR estimated from the separation model output falls below 0~dB, as quality estimation for weak sEMG signals at low SNR is more vulnerable to NN-induced artifacts. In contrast, applying CFV at high SNR provides limited benefit and may inadvertently remove genuine physiological components with low power.

\begin{figure}[t!]
    \centering
    \includegraphics[width=.85\columnwidth]{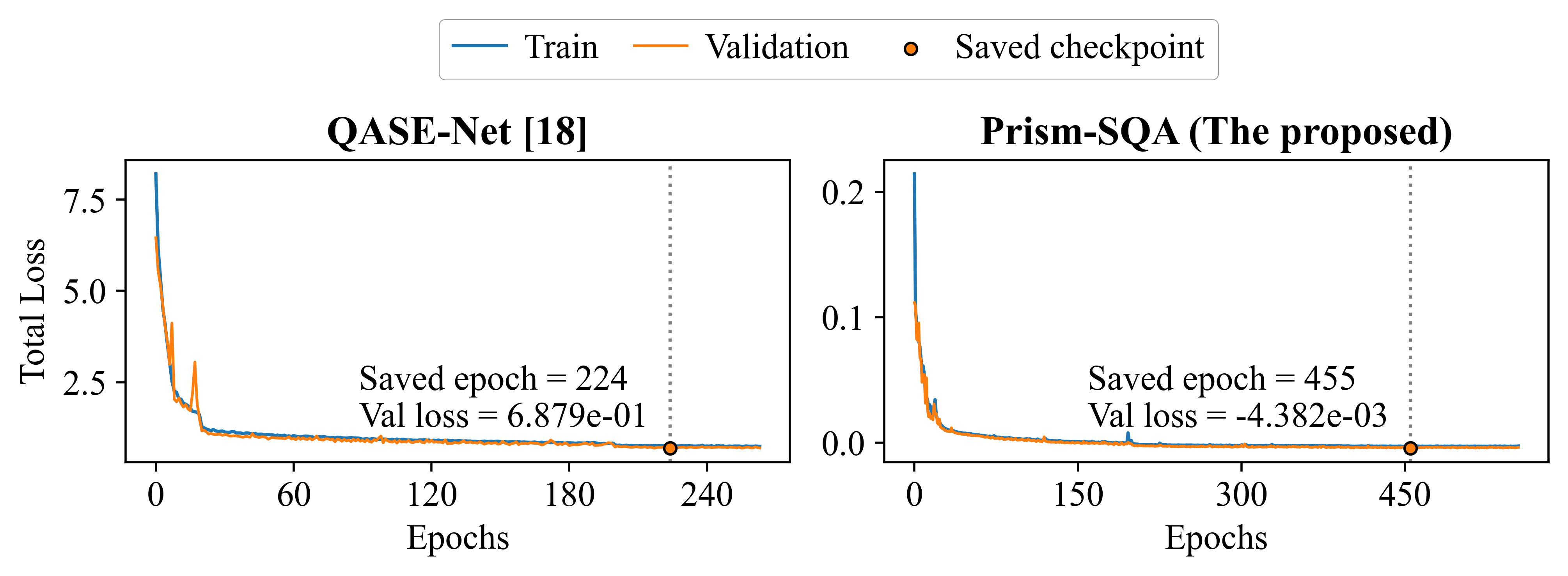}
    \caption{Training and validation loss of Prism-SQA and QASE-Net.}
    \label{fig:loss_curve}
\end{figure}

\subsubsection{Quality index extraction and customization}
In the final stage, quality indices are extracted from the CFV-validated components, providing the proposed framework with adaptability to diverse sEMG applications. Specifically, we extract six indices: the overall SNR and five contaminant-specific SNRs—SPR, SLR, SER, SSR, and SHR—defined according to Eq. \eqref{eq:SNR}. Since these indices are derived from the CFV-validated components rather than from a learned latent representation, users can flexibly redefine or extend them without retraining the NN. For example, one may exclude contaminants with negligible impact on downstream task performance~\cite{zhao2024biosignal, sauer2024signal} or modify the SNR formulation by using different components or replacing signal power with alternative statistical measures, such as amplitude or variance~\cite{sauer2024signal}.

\subsection{Implementation details}
The separation model was trained using the Adam optimizer~\cite{kingma2015adam} with the objective function defined in Eq. (\ref{eq:total_loss}). The loss weights \(\lambda_1\), \(\lambda_2\), and \(\lambda_3\) were set to 1, 0.001, and 0.001, respectively. Training was performed with a batch size of 256 and an initial learning rate of 0.01, which was reduced by a factor of 10 after 20, 200, and 400 epochs using a learning rate scheduler. An early-stopping strategy was applied to prevent overfitting, stopping training when the validation loss did not improve for 100 consecutive epochs, while preserving the model parameters with the lowest validation loss. The training and validation loss curves of Prism-SQA and QASE-Net are shown in Fig. \ref{fig:loss_curve}, demonstrating stable convergence behavior for both models during training. All experiments were implemented in Python (version 3.10) and the PyTorch library (version 2.1.0) on Linux (Ubuntu 22.04.4 LTS) and executed on a single NVIDIA GeForce RTX 3090 GPU.

\subsection{Performance evaluation}

\subsubsection{Evaluation metrics}
\label{para: evaluation metric}
    Different evaluation criteria were used for the two experiments. In the first experiment, the goal was to estimate the signal quality score of each input sEMG segment, for which ground-truth values were available. Performance was assessed using three metrics that quantify the correlation and error between estimated and ground-truth scores: the linear correlation coefficient (LCC), Spearman’s rank correlation coefficient (SRCC)~\cite{spearman1961general}, and mean absolute error (MAE). These metrics are commonly used to assess the accuracy of signal quality estimation models~\cite{lee2024non,sauer2024signal}.  
    \textcolor{black}{LCC, SRCC, and MAE measure linear agreement, rank-order consistency, and average absolute error, respectively. Higher LCC and SRCC and lower MAE indicate better performance.} For the second experiment, we used the area under the receiver operating characteristic curve (AUC), since the task involves binary quality classification. AUC represents the probability that a randomly chosen good-quality segment receives a higher estimated score than a randomly chosen poor-quality segment, providing a threshold-independent measure of discriminative capability.
    
    \textcolor{black}{For statistical testing, performance comparisons and ablation analyses in the first experiment were performed at the subject level using paired two-sided \textit{t}-tests. Effect estimates were reported as paired mean differences ($\Delta$, primary method/configuration minus comparator) with 95\% confidence intervals (CIs). The Benjamini--Hochberg procedure was applied separately to three comparison families to control the false discovery rate at 0.05: the SQA performance comparison (126 comparisons), loss ablation (14 comparisons), and CFV ablation (7 comparisons). All reported \textit{p}-values for these analyses are adjusted. For the second experiment, subject-aware 95\% CIs of the AUCs were estimated using a subject-cluster percentile bootstrap with 10,000 resamples to account for within-subject clustering.}

\subsubsection{Baseline SQA approach}

To evaluate the performance of Prism-SQA, we adopted QASE-Net~\cite{lee2024non}, a representative neural approach for sEMG quality quantification, as the baseline. This comparison aims to examine whether Prism-SQA can achieve competitive quality estimation performance while providing superior interpretability and adaptability, which are generally lacking in existing black-box designs. 
For a fair comparison, QASE-Net’s model capacity was scaled to match Prism-SQA by increasing the number of filters in its convolutional layers and the hidden dimensions of its BiLSTM, attention layer, and MLP, resulting in approximately 19 million parameters (compared with 16 million for Prism-SQA). Its final linear layer was also modified to predict the same six quality indices as Prism-SQA. A detailed comparison of the computational complexity, including the number of parameters, FLOPs, and inference time, is presented in Section \ref{subsect:limitations}.

For the experiment on quality classification, we additionally included a conventional feature-based SQA method as a baseline alongside NN-based approaches. This baseline adopts the variance of the sEMG waveform as the quality index, motivated by the widespread use of dispersion-based measures in signal quality assessment~\cite{del2011assessment,cuadros2022automatic}. Moreover, variance-based SQA has shown strong discriminative performance on the SQI dataset, achieving an AUC of 0.9 on SQI-DB1 in the original study~\cite{cuadros2022automatic}. This comparison enables benchmarking Prism-SQA against both NN-based and handcrafted-feature SQA methods under the same evaluation setting.

\begin{figure*}[!t]
    \centering
    \subfloat[Overall scenario]{
        \includegraphics[width=0.42\textwidth]
        {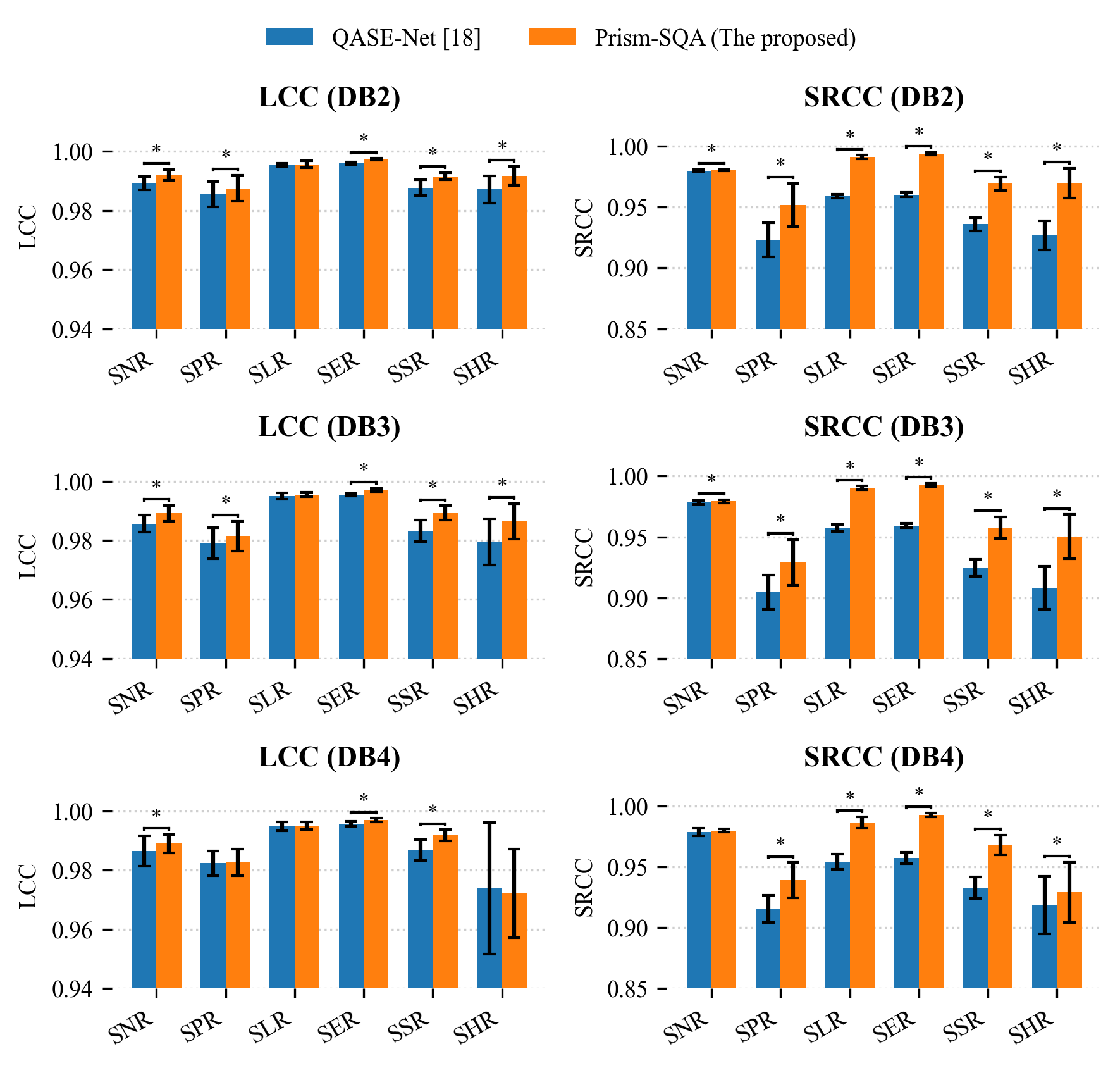}
        \label{fig:overall_CC}
    }
    \hspace{0.015\textwidth}
    \subfloat[Contaminant-included scenario]{
        \includegraphics[width=0.42\textwidth]
        {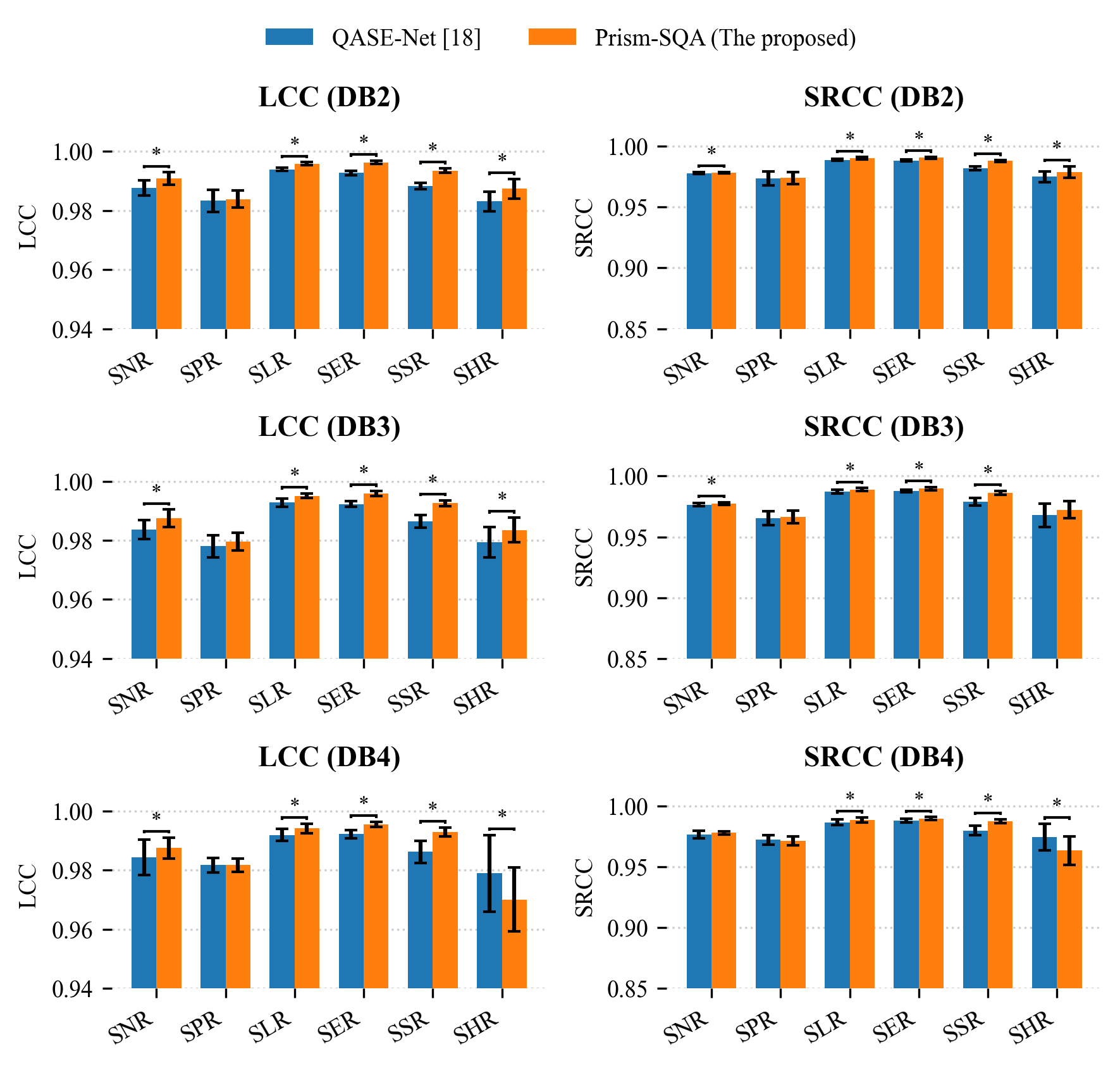}
        \label{fig:Noise_included_CC}
    }

    \caption{\textcolor{black}{Correlation between the estimated and ground-truth quality scores across three datasets (DB2--DB4) under (a) overall and (b) contaminant-included scenarios. Rows correspond to DB2--DB4, and columns show LCC and SRCC, respectively. Bars and error bars represent the mean and standard deviation across subjects for six quality indices (SNR, SPR, SLR, SER, SSR, and SHR). Asterisks indicate statistically significant differences between Prism-SQA and QASE-Net (\textit{p}-value $<$ 0.05).}}
    \label{fig:corr_combined}
\end{figure*}

\renewcommand{\arraystretch}{1}
\begin{table*}[t]
\scriptsize
\centering
\caption{Performance of two SQA methods in terms of MAE (unit: dB) for DB2--DB4.}
\begin{tabular}{cccccccc}
\toprule
 & & \multicolumn{2}{c}{Overall} & \multicolumn{2}{c}{Contaminant-included} & \multicolumn{2}{c}{Contaminant-excluded} \\
\cmidrule(lr){3-4} \cmidrule(lr){5-6} \cmidrule(lr){7-8}
Dataset & Quality index & QASE-Net \cite{lee2024non} & The proposed & QASE-Net \cite{lee2024non} & The proposed & QASE-Net \cite{lee2024non} & The proposed \\
\midrule
& SNR & 1.316 $\pm$ 0.120 &
\textbf{\textcolor{black}{0.948 $\pm$ 0.105}}* &
1.363 $\pm$ 0.123 &
\textbf{\textcolor{black}{0.984 $\pm$ 0.108}}* &
0.145 $\pm$ 0.080 &
\textbf{0.063 $\pm$ 0.088}* \\

& SPR & 1.299 $\pm$ 0.222 &
\textbf{\textcolor{black}{1.161 $\pm$ 0.253}}* &
1.727 $\pm$ 0.230 &
\textbf{\textcolor{black}{1.665 $\pm$ 0.273}}\phantom{*} &
0.707 $\pm$ 0.238 &
\textbf{0.466 $\pm$ 0.229}* \\

DB2 & SLR & 0.732 $\pm$ 0.035 &
\textbf{\textcolor{black}{0.547 $\pm$ 0.042}}* &
1.128 $\pm$ 0.053 &
\textbf{0.811 $\pm$ 0.049}* &
\textbf{0.177 $\pm$ 0.020} &
\textbf{\textcolor{black}{0.177 $\pm$ 0.043}}\phantom{*} \\

& SER & 0.692 $\pm$ 0.032 &
\textbf{\textcolor{black}{0.484 $\pm$ 0.027}}* &
1.095 $\pm$ 0.044 &
\textbf{0.772 $\pm$ 0.037}* &
0.132 $\pm$ 0.025 &
\textbf{\textcolor{black}{0.084 $\pm$ 0.022}}* \\

& SSR & 1.154 $\pm$ 0.128 &
\textbf{\textcolor{black}{0.973 $\pm$ 0.093}}* &
1.464 $\pm$ 0.079 &
\textbf{1.068 $\pm$ 0.050}* &
\textbf{0.729 $\pm$ 0.191} &
\textcolor{black}{0.843 $\pm$ 0.157}* \\
 & SHR & 1.268 $\pm$ 0.218 & \textbf{1.052 $\pm$ 0.183}* & 1.793 $\pm$ 0.177 & \textbf{1.628 $\pm$ 0.204}* & 0.545 $\pm$ 0.323 & \textbf{0.255 $\pm$ 0.154}* \\
\midrule
& SNR & 1.502 ± 0.127 & \textbf{1.114 ± 0.131}* & 1.554 ± 0.130 & \textbf{1.153 ± 0.134}* & 0.207 ± 0.151 & \textbf{0.140 ± 0.103}\phantom{*} \\
     & SPR & 1.609 ± 0.209 & \textbf{1.467 ± 0.261}* & 2.039 ± 0.202 & \textbf{2.000 ± 0.323}\phantom{*} & 1.016 ± 0.335 & \textbf{0.734 ± 0.304}* \\
DB3  & SLR & 0.796 ± 0.072 & \textbf{0.599 ± 0.057}* & 1.235 ± 0.106 & \textbf{0.897 ± 0.084}* & \textbf{0.179 ± 0.039} & 0.180 ± 0.033\phantom{*} \\
     & SER & 0.760 ± 0.059 & \textbf{0.539 ± 0.045}* & 1.176 ± 0.086 & \textbf{0.831 ± 0.066}* & 0.174 ± 0.031 & \textbf{0.126 ± 0.027}* \\
     & SSR & 1.381 ± 0.143 & \textbf{1.134 ± 0.151}* & 1.613 ± 0.117 & \textbf{1.152 ± 0.087}* & \textbf{1.060 ± 0.284} & 1.109 ± 0.293\phantom{*} \\
     & SHR & 1.631 ± 0.314 & \textbf{1.368 ± 0.375}* & 2.044 ± 0.417 & \textbf{1.993 ± 0.514}\phantom{*} & 1.061 ± 0.593 & \textbf{0.501 ± 0.259}* \\
\midrule
& SNR & 1.473 ± 0.232 & \textbf{1.184 ± 0.160}* & 1.522 ± 0.238 & \textbf{1.221 ± 0.165}* & \textbf{0.264 ± 0.138} & 0.269 ± 0.236\phantom{*} \\
     & SPR & 1.437 ± 0.187 & \textbf{1.352 ± 0.210}* & 1.791 ± 0.175 & \textbf{1.734 ± 0.171}* & 0.955 ± 0.304 & \textbf{0.831 ± 0.308}* \\
DB4   & SLR & 0.829 ± 0.091 & \textbf{0.632 ± 0.069}* & 1.294 ± 0.150 & \textbf{0.965 ± 0.108}* & 0.179 ± 0.033 & \textbf{0.167 ± 0.031}\phantom{*} \\
     & SER & 0.764 ± 0.066 & \textbf{0.547 ± 0.041}* & 1.204 ± 0.110 & \textbf{0.867 ± 0.064}* & 0.151 ± 0.021 & \textbf{0.102 ± 0.017}* \\
     & SSR & 1.246 ± 0.165 & \textbf{0.974 ± 0.144}* & 1.566 ± 0.202 & \textbf{1.125 ± 0.098}* & 0.807 ± 0.185 & \textbf{0.768 ± 0.216}\phantom{*} \\
     & SHR & \textbf{1.750 ± 0.676} & 1.865 ± 0.490\phantom{*} & \textbf{1.779 ± 0.349} & 2.156 ± 0.317* & 1.708 ± 1.154 & \textbf{1.457 ± 0.761}\phantom{*} \\
     \bottomrule
\multicolumn{8}{l}{\scriptsize *Indicates a statistically significant difference (\textit{p}-value $<$ 0.05) between Prism-SQA and QASE-Net. \textbf{Bold} font indicates the better score for each quality index.}
\end{tabular}
\label{tab:result_MAE}
\end{table*}

\begin{figure*}[t!]
    \centering
    \includegraphics[width=.8\textwidth]{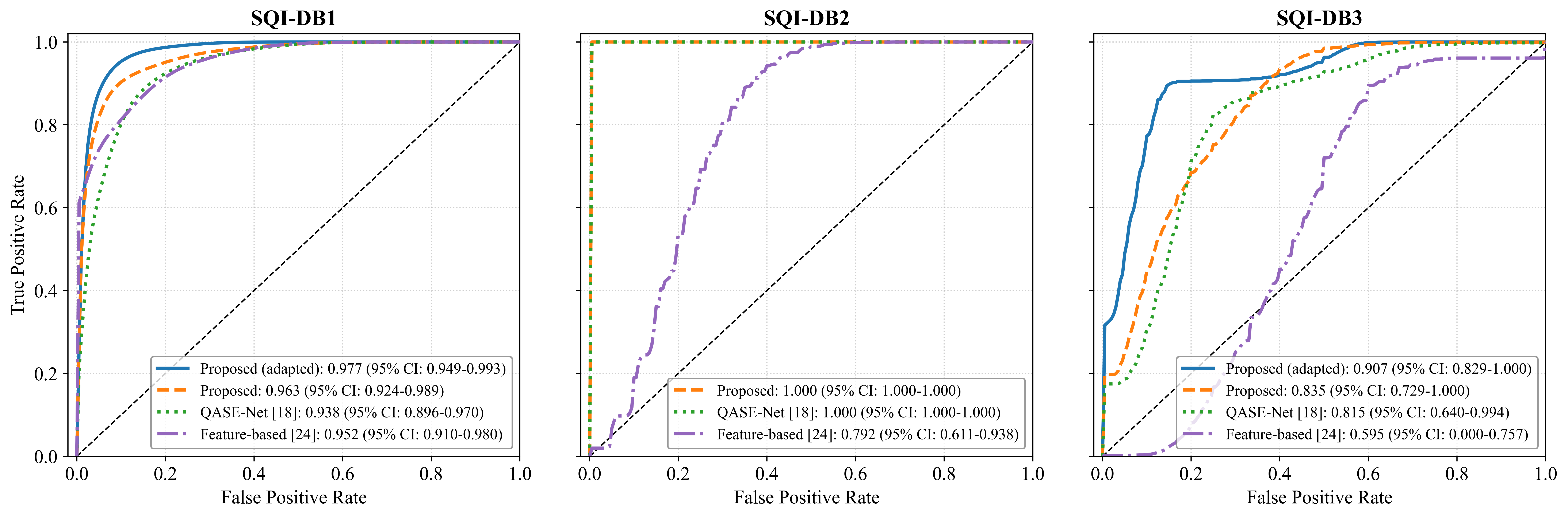}
    \caption{\textcolor{black}{ROC curves and corresponding AUCs with 95\% CIs for signal quality classification on the SQI-DB1--DB3 datasets. The adapted Prism-SQA results for SQI-DB1 and SQI-DB3 are reported as an exploratory demonstration of quality-index adaptability.}}
    \label{fig:result_AUC}
\end{figure*}

\begin{table*}[t]
\centering
\scriptsize
\caption{Ablation on the loss terms of the objective function.}
\resizebox{\textwidth}{!}{%
\begin{tabular}{cccccccccc}
\toprule
\multicolumn{3}{c}{Components} & \multicolumn{3}{c}{Overall} & \multicolumn{3}{c}{Contaminant-included} & \multicolumn{1}{c}{Contaminant-excluded} \\
\cmidrule(lr){1-3} \cmidrule(lr){4-6} \cmidrule(lr){7-9} \cmidrule(lr){10-10}
$\mathcal{L}_{1}$ & $\mathcal{L}_{\text{SD-SDR}}$  & $\mathcal{L}_{\mathrm{cls}}$ &LCC $\uparrow$ & SRCC $\uparrow$ & MAE $\downarrow$ & LCC $\uparrow$ & SRCC $\uparrow$ & MAE $\downarrow$ & MAE $\downarrow$ \\
\midrule
\checkmark & \checkmark & \checkmark & \textbf{0.988 $\pm$ 0.004}\phantom{*} & \textbf{0.965 $\pm$ 0.009}\phantom{*} & \textbf{1.059 $\pm$ 0.179}\phantom{*} & 0.989 $\pm$ 0.003\phantom{*} & \textbf{0.981 $\pm$ 0.003}\phantom{*} & 1.276 $\pm$ 0.175\phantom{*} & 0.613 $\pm$ 0.192\phantom{*} \\
\checkmark & \checkmark & – & 0.986 $\pm$ 0.004* & 0.961 $\pm$ 0.010* & 1.161 $\pm$ 0.228* & \textbf{0.990 $\pm$ 0.002}\phantom{*} & 0.978 $\pm$ 0.006* & \textbf{1.166 $\pm$ 0.207}* & 1.060 $\pm$ 0.238* \\
\checkmark & – & – & 0.981 $\pm$ 0.004* & 0.944 $\pm$ 0.009* & 1.816 $\pm$ 0.230* & 0.987 $\pm$ 0.003* & 0.977 $\pm$ 0.003* & 2.513 $\pm$ 0.345* & \textbf{0.594 $\pm$ 0.087}\phantom{*} \\
\bottomrule
\multicolumn{10}{l}{\scriptsize *Denotes a significant difference (\textit{p}-value $<$ 0.05) from the full loss configuration. \textbf{Bold} font indicates the best score for each metric. } \\
\multicolumn{10}{l}{\scriptsize$\uparrow$ and $\downarrow$ denote that better performance is achieved with higher and lower metric values, respectively.}
\end{tabular}
}
\label{tab:ablation_loss}
\end{table*}

\begin{table*}[t]
\scriptsize
\centering
\caption{Ablation on the CFV module of the proposed framework.}
\begin{tabular}{cccccccc}
\toprule
\multicolumn{1}{c}{} & \multicolumn{3}{c}{Overall} & \multicolumn{3}{c}{Contaminant-included} & \multicolumn{1}{c}{Contaminant-excluded} \\
\cmidrule(lr){2-4} \cmidrule(lr){5-7} \cmidrule(lr){8-8}
Variant & LCC $\uparrow$ & SRCC $\uparrow$ & MAE $\downarrow$ & LCC $\uparrow$ & SRCC $\uparrow$ & MAE $\downarrow$ & MAE $\downarrow$ \\
\midrule
w CFV
& \textbf{\textcolor{black}{0.9901 $\pm$ 0.0034}}\phantom{*}
& \textbf{\textcolor{black}{0.9693 $\pm$ 0.0085}}\phantom{*}
& \textbf{0.995 $\pm$ 0.175}\phantom{*}
& \textcolor{black}{0.9891 $\pm$ 0.0028}\phantom{*}
& \textcolor{black}{0.9810 $\pm$ 0.0031}\phantom{*}
& 1.277 $\pm$ 0.174\phantom{*}
& \textbf{0.458 $\pm$ 0.185}\phantom{*} \\

w/o CFV
& \textcolor{black}{0.9881 $\pm$ 0.0035}*
& \textcolor{black}{0.9651 $\pm$ 0.0087}*
& 1.059 $\pm$ 0.179*
& \textbf{\textcolor{black}{0.9893 $\pm$ 0.0029}}\textcolor{black}{*}
& \textbf{\textcolor{black}{0.9812 $\pm$ 0.0031}}\textcolor{black}{*}
& \textbf{1.276 $\pm$ 0.175}*
& 0.613 $\pm$ 0.192* \\
\bottomrule
\multicolumn{8}{l}{\scriptsize *Denotes a significant difference (\textit{p}-value $<$ 0.05) from the proposed method. \textbf{Bold} font indicates the best score for each metric.} \\
\multicolumn{8}{l}{$\uparrow$ and $\downarrow$ denote that better performance is achieved with higher and lower metric values, respectively.}
\end{tabular}
\label{tab:ablation_CFV}
\end{table*}

\section{Results}
\label{sec:result}

We first present the results of quality score estimation using the proposed Prism-SQA, evaluated in terms of LCC, SRCC, and MAE, on the test sets derived from the three Ninapro sub-databases (DB2, DB3, and DB4) in Subsection~\ref{subsec: result_quality_score}. Subsection~\ref{subsec: result_quality_classification} then reports the quality classification performance (AUC) on the clinical SQI dataset, \textcolor{black}{with the unadapted Prism-SQA treated as the primary zero-shot clinical evaluation and the adapted SNR reported as an exploratory demonstration of quality-index adaptability.} Finally, Subsection~\ref{subsec:result_ablation} provides the ablation study on the design of Prism-SQA, including the effects of the objective function and the CFV module. \textcolor{black}{All NN models are evaluated without fine-tuning.}

\subsection{Quality score estimation}
\label{subsec: result_quality_score}
We evaluated the performance of quality-score estimation under three complementary scenarios—overall, contaminant-included, and contaminant-excluded—to assess the robustness of the proposed SQA framework. These scenarios reflect the two fundamental requirements of SQA: accurately quantifying noise severity when contamination is present and avoiding false detections when the contamination is absent. The overall evaluation considers all test samples and summarizes model performance across diverse noise conditions.
The contaminant-included evaluation focuses on samples containing the contaminant associated with the assessed quality index (e.g., sEMG with LFN for SLR), allowing for an assessment of how well the model estimates contaminant severity. For Prism-SQA, this also reflects reconstruction fidelity for relevant components.
The contaminant-excluded evaluation examines samples where the relevant contaminant is absent (e.g., sEMG without LFN for SLR\footnote{For the general SNR index, contaminant-included and -excluded scenarios correspond to noisy versus clean sEMG segments.}), assessing whether the model avoids spurious estimates of the contaminant. This is also an indicator of Prism-SQA’s ability to generate zero vectors with no artifacts for the absent contaminants. Because contaminant-excluded segments have a constant ground-truth score (30 dB in this study), correlation metrics are not meaningful and are therefore omitted. All reported results are presented as means and standard deviations across subjects within each dataset.

Fig.~\ref{fig:overall_CC} presents the overall correlation performance, in terms of LCC and SRCC, for the proposed Prism-SQA and the baseline QASE-Net across Ninapro DB2--DB4 test sets. Both methods achieve consistently high correlations, with LCCs exceeding 0.97 and SRCCs above 0.90 across all six quality indices. Prism-SQA achieves higher correlation scores in most cases, with statistically significant improvements (\textit{p}-value $<$ 0.05) in the majority of comparisons. Specifically, for LCC, Prism-SQA performs significantly better in five, \textcolor{black}{five}, and three of the six quality indices on DB2, DB3, and DB4, respectively. In the remaining cases, no significant difference is observed. For SRCC, Prism-SQA outperforms QASE-Net in six, six, and five indices on DB2, DB3, and DB4, respectively. For both LCC and SRCC, a slight degradation is observed as the test domain shifts from DB2 to DB3 and DB4, suggesting that cross-dataset differences introduce mild performance drops due to domain variability.

Fig.~\ref{fig:Noise_included_CC} presents the correlation performance of Prism-SQA and QASE-Net under the contaminant-included scenario. Both methods maintain strong correlations, achieving average LCC and SRCC values above 0.96 across all quality indices. Prism-SQA \textcolor{black}{generally} attains higher correlations than QASE-Net, with statistically significant improvements (\textit{p}-value $<$ 0.05) in most cases. Specifically, for LCC, Prism-SQA significantly outperforms QASE-Net in five, five, and four of the six quality indices on DB2, DB3, and DB4, respectively. \textcolor{black}{QASE-Net significantly outperforms Prism-SQA only for SHR in DB4 ($\Delta\mathrm{LCC}=-0.009$; 95\% CI [$-0.014$, $-0.004$]; $p=5.49\times10^{-3}$), while the remaining cases show no significant differences.} For SRCC, Prism-SQA shows significant improvements in five, four, and three indices on DB2, DB3, and DB4, respectively. \textcolor{black}{Similarly, QASE-Net significantly outperforms Prism-SQA only for SHR in DB4 ($\Delta\mathrm{SRCC}=-0.011$; 95\% CI [$-0.016$, $-0.006$]; $p=7.77\times10^{-4}$), while the remaining cases show no significant differences.}

Table~\ref{tab:result_MAE} summarizes the error performance across the three scenarios and datasets.
Prism-SQA consistently achieves lower MAE than QASE-Net under most conditions, with differences being statistically significant (\textit{p}-value $<$ 0.05).
Under the overall scenario, Prism-SQA yields significantly lower MAE under all conditions except for the SHR index on DB4, where no significant difference is found between the two methods \textcolor{black}{($\Delta\mathrm{MAE}=0.115$~dB; 95\% CI [$-0.081$, $0.310$]; $p=0.248$)}.
\textcolor{black}{Under the contaminant-included scenario, Prism-SQA yields significantly lower MAE in 14 of the 18 comparisons. No significant differences are observed for SPR in DB2 and DB3 and SHR in DB3, whereas QASE-Net achieves significantly lower MAE only for SHR in DB4 ($\Delta\mathrm{MAE}=0.377$~dB; 95\% CI [$0.225$, $0.529$]; $p=5.08\times10^{-4}$).}
In the contaminant-excluded scenario, Prism-SQA exhibits significantly lower MAE in 9 out of 18 cases (6 quality indices $\times$ 3 datasets), including SPR and SER across all datasets, SNR in DB2, and SHR in DB2 and DB3. QASE-Net only exhibits significantly lower MAE for one case, SSR in DB2 \textcolor{black}{($\Delta\mathrm{MAE}=0.115$~dB; 95\% CI [$0.060$, $0.169$]; $p=1.38\times10^{-3}$)}. The remaining eight cases show no significant difference between the two methods.

\begin{figure}[t!]
    \centering
    \includegraphics[width=.7\columnwidth]{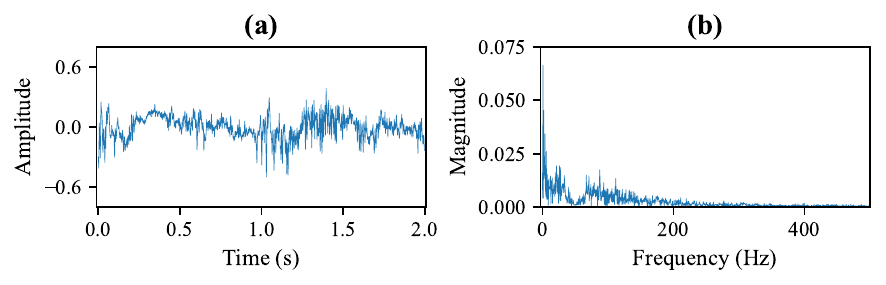}
    \caption{(a) Waveform and (b) frequency spectrum of an sEMG recording from SQI-DB3 annotated as good-quality despite containing LFN.}
    \label{fig:good_quality_lfn_case}
\end{figure}

\subsection{Quality classification}
\label{subsec: result_quality_classification}
For the quality classification task on the clinical SQI dataset, the SNRs estimated by QASE-Net and Prism-SQA were used as decision variables to classify signals into good- and poor-quality segments. For Prism-SQA, we additionally report an adapted SNR for SQI-DB1 and SQI-DB3 \textcolor{black}{as an exploratory demonstration of its adaptability to dataset-specific quality criteria}. This adaptation was motivated by the annotation characteristics of SQI-DB1 and SQI-DB3, where LFN-containing sEMG segments can still be labeled as good-quality data (shown in Fig.~\ref{fig:good_quality_lfn_case}). \textcolor{black}{Accordingly, the adapted SNR excludes the LFN component from the noise-power term, in contrast to the conventional SNR, which treats all separated contaminant components as noise. The adapted quality index is defined as:}

\begin{equation}
\small
SNR_{\text{adapted}} = 10\log_{10}
\left(
\frac{P_{\text{sEMG}}}
{P_{\text{PLI}} + P_{\text{ECG}} + P_{\text{SPK}} + P_{\text{HFN}}}
\right),
\end{equation}
where $P_x$ denotes the power of the separated component $x$.

Fig.~\ref{fig:result_AUC} presents the classification performance in terms of ROC curves and the corresponding AUCs with 95\% CIs. \textcolor{black}{In the primary zero-shot evaluation,} Prism-SQA (orange) achieves AUCs of 0.963 \textcolor{black}{(0.924--0.989)}, 1.000 \textcolor{black}{(1.000--1.000)}, and 0.835 \textcolor{black}{(0.729--1.000)} on SQI-DB1, SQI-DB2, and SQI-DB3, respectively.
The corresponding AUCs of QASE-Net (green) are 0.938 \textcolor{black}{(0.896--0.970)}, 1.000 \textcolor{black}{(1.000--1.000)}, and 0.815 \textcolor{black}{(0.640--0.994)}.
\textcolor{black}{Thus, Prism-SQA achieves numerically higher AUCs on SQI-DB1 and SQI-DB3 and identical performance on SQI-DB2, while the overlapping confidence intervals do not establish a statistically supported difference between the two NN-based SQA methods.} The conventional feature-based SQA method (purple) performs competitively on SQI-DB1 with an AUC of 0.952 \textcolor{black}{(0.910--0.980)}, but its performance drops substantially on SQI-DB2 and SQI-DB3, yielding AUCs of 0.792 \textcolor{black}{(0.611--0.938)}\footnote{The feature-based method behaves differently on SQI-DB2, where higher variance indicates better quality, yet we report its best achievable performance.} and 0.595 \textcolor{black}{(0.000--0.757)}, respectively.

\textcolor{black}{In the exploratory adaptability analysis, the adapted Prism-SQA (blue) yields AUCs of 0.977 (0.949--0.993) and 0.907 (0.829--1.000) on SQI-DB1 and SQI-DB3, respectively, which are numerically higher than those of the unadapted version. No adapted result is reported for SQI-DB2 because the LFN exclusion was motivated specifically by the annotation characteristics of SQI-DB1 and SQI-DB3.}

\subsection{Ablation study}
\label{subsec:result_ablation}
To assess the contribution of each design component in the proposed SQA framework, we conducted ablation studies on the objective function and the CFV module. The experiment evaluated the performance in quality score estimation using the test sets derived from Ninapro DB2, DB3, and DB4. For each scenario, the average values of all six quality indices (SNR, SPR, SLR, SER, SSR, and SHR) were calculated for each subject, and the final results were obtained by aggregating across all subjects from all three datasets.

\subsubsection{Ablation study on the objective function}
To assess the contribution of each loss term, we conducted a progressive ablation of the proposed objective function with CFV disabled for all configurations. The configuration ($\mathcal{L}_{1}$ + $\mathcal{L}_{\text{SD-SDR}}$ + $\mathcal{L}_{\mathrm{cls}}$) represents the proposed loss function setting, while two reduced variants were evaluated: ($\mathcal{L}_{1}$ + $\mathcal{L}_{\text{SD-SDR}}$) and ($\mathcal{L}_{1}$). The results across the three test scenarios are summarized in Table~\ref{tab:ablation_loss}.

\textcolor{black}{Defining $\Delta$ as the paired difference between the full and reduced configurations, comparison with $\mathcal{L}_{1}$ alone shows overall improvements of $\Delta\mathrm{LCC}=0.007$ (95\% CI [$0.006$, $0.007$]; $p=1.80\times10^{-20}$), $\Delta\mathrm{SRCC}=0.022$ ([$0.021$, $0.022$]; $p=6.93\times10^{-34}$), and $\Delta\mathrm{MAE}=-0.757$~dB ([$-0.808$, $-0.706$]; $p=2.38\times10^{-23}$).} When only $\mathcal{L}_{1}$ is used, both correlation and error metrics deteriorate, with LCC and SRCC dropping to 0.981 and 0.944, and MAE increasing to 1.816~dB, mainly due to degraded performance under contaminant-included conditions. Incorporating $\mathcal{L}_{\text{SD-SDR}}$ alongside $\mathcal{L}_{1}$ substantially improves these results (LCC~=~0.986, SRCC~=~0.961, MAE~=~1.161~dB).

\textcolor{black}{Comparison with $\mathcal{L}_{1}+\mathcal{L}_{\text{SD-SDR}}$ further shows that adding $\mathcal{L}_{\mathrm{cls}}$ improves the overall LCC by $\Delta\mathrm{LCC}=0.002$ (95\% CI [$0.001$, $0.002$]; $p=4.56\times10^{-6}$), SRCC by $\Delta\mathrm{SRCC}=0.005$ ([$0.003$, $0.006$]; $p=1.57\times10^{-5}$), and MAE by $\Delta\mathrm{MAE}=-0.102$~dB ([$-0.139$, $-0.064$]; $p=7.13\times10^{-6}$). The MAE reduction is primarily observed under the contaminant-excluded condition, with $\Delta\mathrm{MAE}=-0.447$~dB ([$-0.479$, $-0.415$]; $p=8.87\times10^{-23}$), whereas under the contaminant-included condition, $\mathcal{L}_{\mathrm{cls}}$ increases MAE by $\Delta\mathrm{MAE}=0.110$~dB ([$0.072$, $0.149$]; $p=3.74\times10^{-6}$).}

\subsubsection{Effect of the CFV module}
Table~\ref{tab:ablation_CFV} summarizes the results of the proposed framework with and without the CFV.
\textcolor{black}{Incorporating CFV significantly improves performance in the overall and contaminant-excluded scenarios. Defining $\Delta$ as the paired difference between the configurations with and without CFV, the overall improvements are $\Delta\mathrm{LCC}=0.0021$ (95\% CI [$0.0019$, $0.0022$]; $p=5.51\times10^{-22}$), $\Delta\mathrm{SRCC}=0.0043$ ([$0.0040$, $0.0045$]; $p=5.44\times10^{-25}$), and $\Delta\mathrm{MAE}=-0.064$~dB ([$-0.068$, $-0.060$]; $p=9.00\times10^{-25}$). The largest} \textcolor{black}{MAE reduction is observed under the contaminant-excluded condition, with $\Delta\mathrm{MAE}=-0.155$~dB ([$-0.164$, $-0.146$]; $p=5.44\times10^{-25}$).}
\textcolor{black}{Under the contaminant-included scenario, the differences are statistically significant but numerically small, with $\Delta\mathrm{LCC}=-0.0002$, $\Delta\mathrm{SRCC}=-0.0002$, and $\Delta\mathrm{MAE}=0.002$~dB.}

\subsection{Sensitivity analysis of CFV threshold values}
\label{subsec:sensitivity of CFV}
\begin{figure}[t!]
   \centering
   \includegraphics[width=.95\columnwidth]{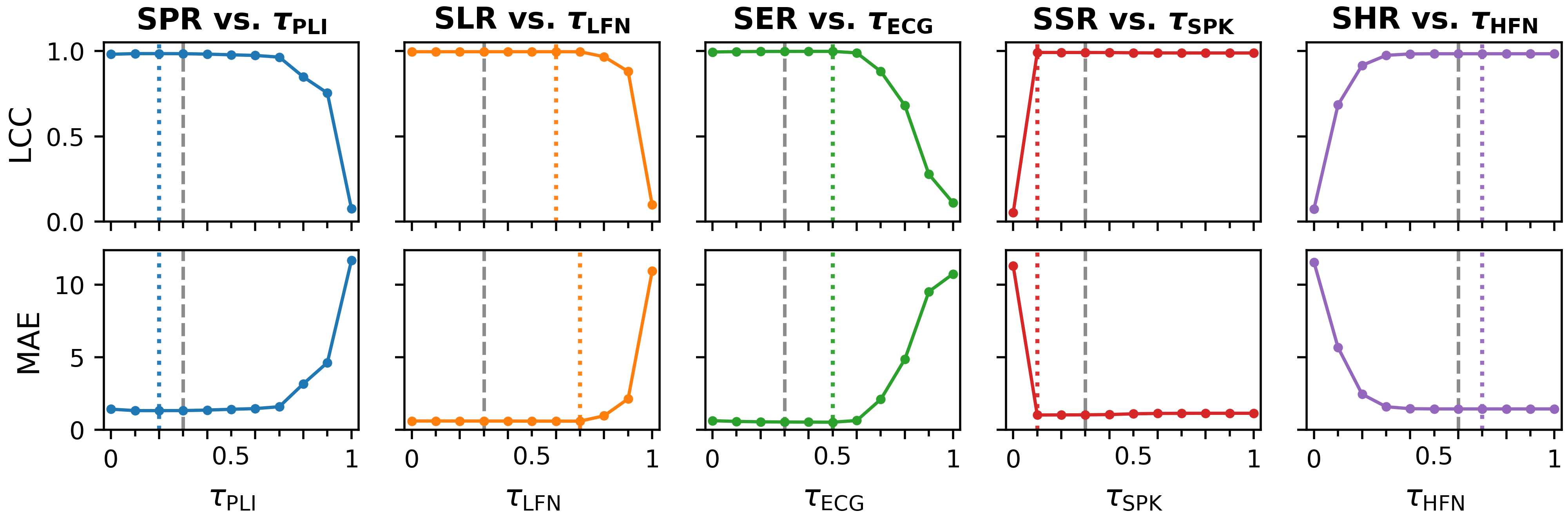}
   \caption{\textcolor{black}{Sensitivity analysis of the CFV thresholds. LCC and MAE of each quality index are evaluated by varying the corresponding threshold. Gray and colored vertical lines denote the adopted and best-performing values, respectively.}}
   \label{fig:CFV sensitivity}
\end{figure}

\textcolor{black}{Fig.~\ref{fig:CFV sensitivity} evaluates the sensitivity of Prism-SQA to the CFV thresholds on Ninapro DB2. Each threshold was varied independently from 0 to 1 while the others were fixed at their default values (e.g., SLR was evaluated by varying only $\tau_{\mathrm{LFN}}$), and the corresponding quality index was evaluated using LCC and MAE. Gray dashed lines indicate the adopted thresholds, and colored dotted lines indicate the best-performing values. Results show that performance remains stable over broad threshold ranges: approximately 0.1--0.6 for the spectral-power-ratio thresholds ($\tau_{\mathrm{PLI}}$, $\tau_{\mathrm{LFN}}$, $\tau_{\mathrm{ECG}}$, and $\tau_{\mathrm{SPK}}$) and 0.4--1.0 for the kurtosis threshold $\tau_{\mathrm{HFN}}$. Although not always optimal, the selected thresholds lie within these stable regions, indicating limited sensitivity to their exact values.}

\section{Discussion}
\label{sec:discussion}

\subsection{Quality estimation performance of Prism-SQA}

Across both experiments, the results highlight how Prism-SQA’s physiology-aware decomposition and verification strategy effectively mitigates the limitations of existing black-box SQA approaches. By separating sEMG into distinct contaminant components and validating each through canonical spectral and temporal fingerprints, Prism-SQA preserves neuromuscular information that end-to-end models often obscure. This leads to more stable and meaningful quality estimates across varying subject populations, contaminant types, and recording conditions—factors that commonly challenge conventional quality indices and handcrafted features. The CFV module further enhances robustness by enforcing physiologically plausible patterns and suppressing implausible artifacts, improving the reliability of quality interpretation. Importantly, Prism-SQA maintains strong performance when applied to clinically collected sEMG from the SQI dataset, \textcolor{black}{providing preliminary evidence of its potential to generalize beyond controlled synthetic scenarios and support quality assessment under clinical variability.} Collectively, these findings suggest that Prism-SQA provides a \textcolor{black}{promising,} physiologically informed basis for accurate sEMG quality estimation.

The superior performance of Prism-SQA arises from its ability to satisfy two key requirements of SQA-oriented source separation: accurately reconstructing the coarse structure and relative scale of existing components, and suppressing undesired outputs in channels corresponding to absent contaminants. The SD-SDR loss is central to the first requirement, as removing it leads to pronounced degradation under contaminant-included conditions (Table~\ref{tab:ablation_loss}), consistent with prior findings that point-wise reconstruction losses often bias models toward high-energy sources and distort inter-source amplitude relationships \cite{guso2022loss, venkatesh2024real}. To meet the second requirement, Prism-SQA employs a component-presence classification loss that suppresses undesired outputs when contaminants are absent; its effectiveness is reflected in the improved performance under contaminant-excluded settings (Table~\ref{tab:ablation_loss}). This strategy aligns with evidence from audio source separation, where presence-aware supervision improves source separation performance \cite{kong2018joint, karamatli2019audio}. Together, these design principles enable Prism-SQA to deliver reliable quality information by accurately reconstructing existing contaminants while suppressing artifact-like outputs in absent channels, thereby supporting more stable and reliable SQA.

\begin{figure}[t!]
   \centering
   \includegraphics[width=.8\columnwidth]{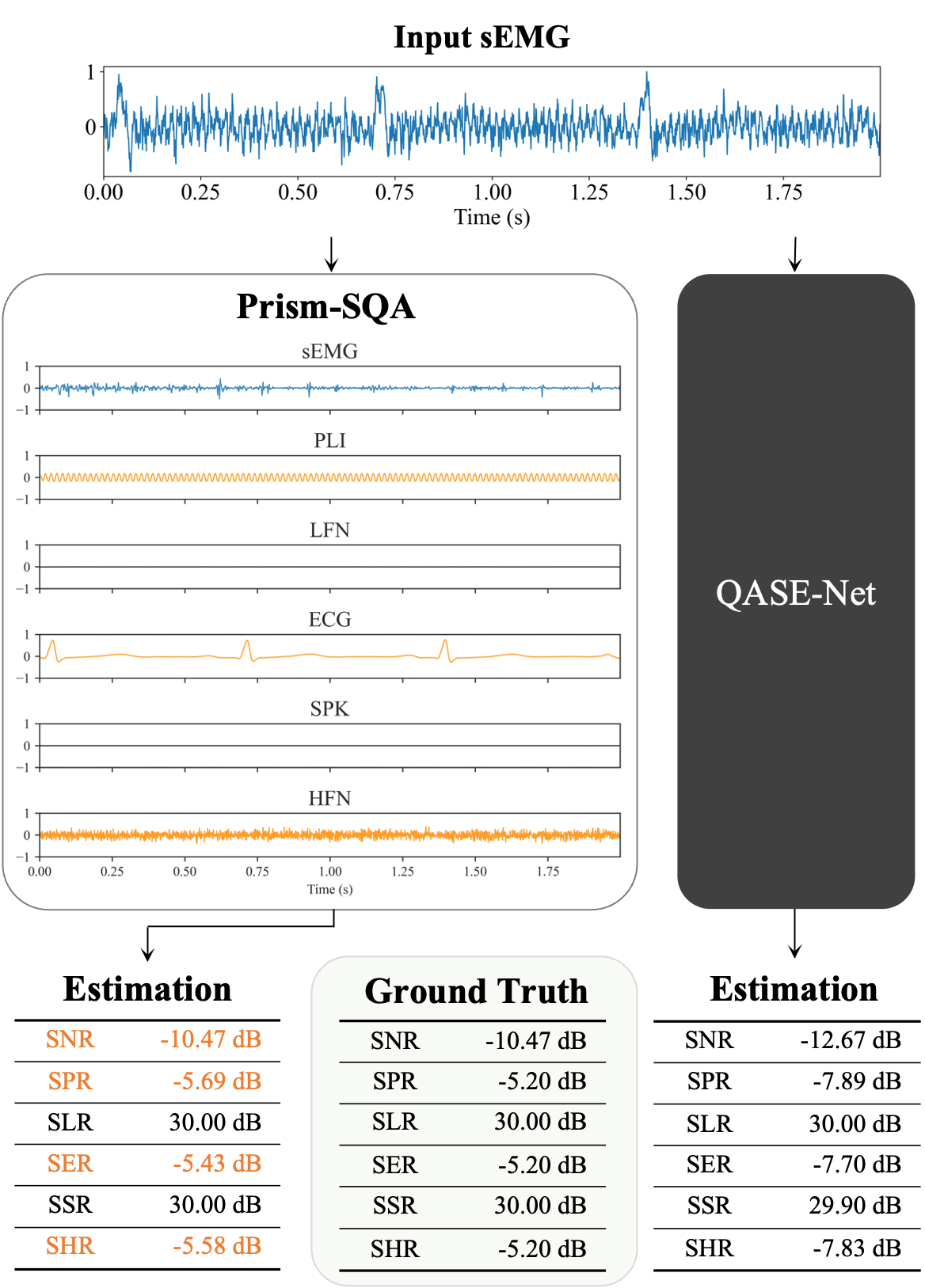}
   \caption{\textcolor{black}{Example illustrating Prism-SQA interpretability. The input sEMG is decomposed into physiologically meaningful components consistent with the actual contamination, while absent contaminants remain zero. Using these components, Prism-SQA generates quality scores that closely match the ground-truth values, providing a transparent, component-level rationale for its estimation. In contrast, black-box models yield quality scores without interpretable representations.}}
   \label{fig:interpretability}
\end{figure}

\begin{figure*}[tbh!]
   \centering
   \includegraphics[width=.7\textwidth]{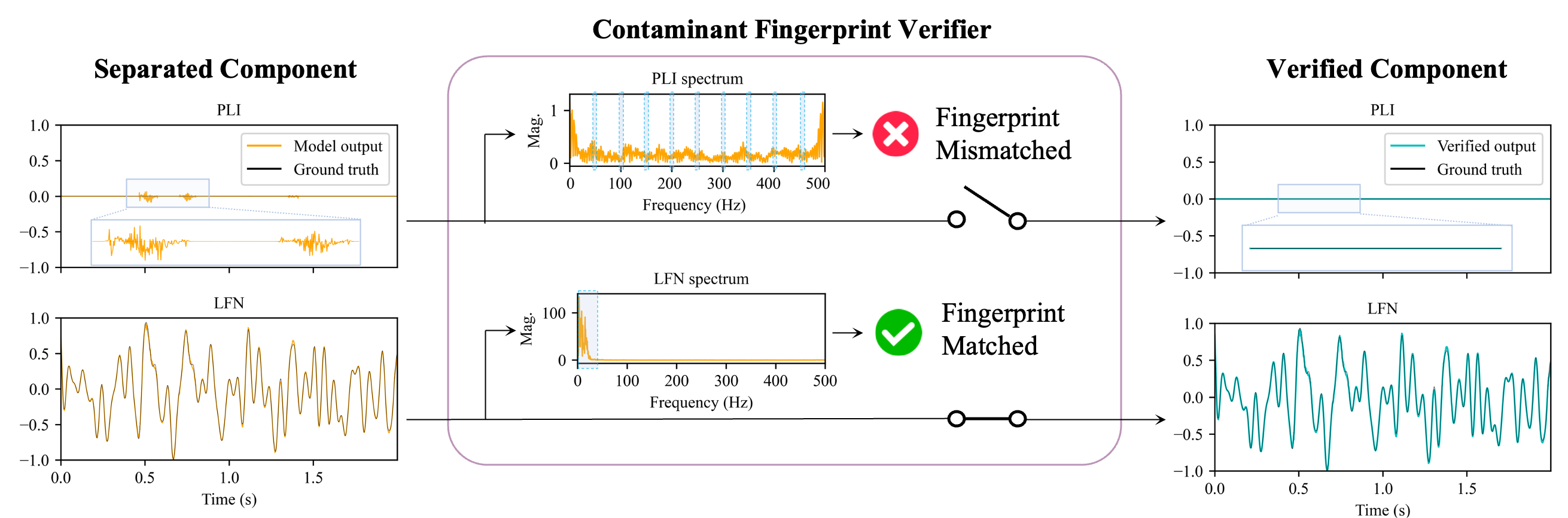}
   \caption{An illustration of artifact rejection by the CFV. The input sEMG is contaminated by LFN at an SNR of $-20$ dB, and the PLI and LFN components are presented. The highlighted artifacts in the PLI component (left) can be discriminated and discarded by the CFV based on the prior knowledge that the component should have a narrowband structure.}
   \label{fig:artifact_waveform}
\end{figure*}

\subsection{Interpretability of Prism-SQA}

Beyond its superior quantitative performance, Prism-SQA provides physiology-consistent and interpretable outputs that enhance the transparency of sEMG SQA. By decomposing each input segment into meaningful components, the model offers a direct, waveform-level rationale for the resulting quality indices. As shown in Fig.~\ref{fig:interpretability}, both SQA approaches produce estimates that approximate the ground truth; however, only Prism-SQA reveals why these estimates are reasonable. The separated PLI, ECG, and HFN components exhibit temporal characteristics consistent with known contaminant patterns, while the LFN and SPK components remain near zero, correctly reflecting the absence of these contaminants. This selective and physiologically grounded decomposition allows each quality index to be traced back to identifiable signal-level evidence, enabling users to apply their domain knowledge to verify whether the estimated quality score is consistent with the underlying contaminant structure. Such component-level interpretability bridges the gap between neural quality estimation and expert electrophysiological reasoning, \textcolor{black}{providing a foundation for more trustworthy neural SQA.}   

Leveraging this interpretability, the CFV module automatically verifies the spectral and statistical plausibility of each separated component.
\textcolor{black}{Ablation results (Table~\ref{tab:ablation_CFV}) show that CFV significantly reduces MAE, particularly under contaminant-excluded conditions, indicating its effectiveness in suppressing spurious contaminant estimates.
Fig.~\ref{fig:artifact_waveform} shows an example where an sEMG segment contaminated only by LFN at an SNR of -20 dB produces spurious activity in the separated PLI component.
CFV rejects this component because it lacks the characteristic narrowband peaks at 50/60 Hz and their harmonics.
Without CFV, the resulting SPR is 13.93 dB, corresponding to a 16.07 dB error from the ground-truth score of 30 dB; after CFV, the error is reduced to 0 dB.
These results demonstrate that CFV can suppress physiologically implausible outputs and improve quality estimation.}

\subsection{Adaptability of Prism-SQA}

Building on the enhanced interpretability, Prism-SQA provides adaptability at the level of quality-index computation, distinguishing it from existing black-box SQA models. By decoupling neural representation learning from analytical index computation, the framework enables users to redefine quality indices without requiring model retraining. \textcolor{black}{This flexibility enables adaptation to different clinical interpretations of signal quality}. \textcolor{black}{For application to a target setting, once the quality criterion has been established based on expert-defined criteria or prior clinical knowledge, users can define the analytical quality index before evaluation.} In the \textcolor{black}{exploratory analysis of the} SQI dataset (Fig. \ref{fig:result_AUC}), applying this adapted quality index resulted in higher AUCs on SQI-DB1 and SQI-DB3, and the resulting quality scores aligned more closely with expert annotations. \textcolor{black}{These results illustrate} the framework’s ability to flexibly accommodate context-specific refinements of conventional quality indices.

More broadly, Prism-SQA's adaptability extends to adopting entirely different definitions of quality indices depending on the application. 
\textcolor{black}{For example, Sauer \textit{et al.} \cite{sauer2024signal} defined respiratory-sEMG quality indices using ECG-to-contaminant ratios and both amplitude- and variance-based measures, which differ substantially from the conventional sEMG-to-contaminant power ratios used in general-purpose SQA \cite{oo2020signal,lee2024non}. Such customized indices can be directly computed from Prism-SQA's separated components by redefining the analytical formulas, without retraining the NN. In contrast, black-box models trained to predict fixed indices, such as QASE-Net, would require retraining for alternative definitions. Although conventional feature-based methods also allow flexible index design, their performance was less robust than NN-based approaches in our clinical evaluation (Fig.~\ref{fig:result_AUC}). Prism-SQA therefore provides a flexible framework for supporting diverse sEMG applications while retaining the performance advantages of NN-based approaches.}

\subsection{Limitations and future research}
\label{subsect:limitations}
This study introduces an interpretable and adaptable neural SQA framework for sEMG, demonstrating competitive performance compared to existing black-box approaches. However, there are several limitations. 
\subsubsection{Generalization to unseen contaminant types and characteristics}
The proposed method relies on a supervised source separation model, which is inherently limited to handling additive contaminant characteristics represented during training. Nonlinear distortions, such as multiplicative WGN \cite{guo2025revisiting}, and atypical contaminant characteristics, such as PLI with uncommon center frequencies observed in certain recording environments \cite{sauer2024signal}, cannot be reliably handled using the current architecture, as they
may not be correctly separated into their corresponding contaminant components and instead remain in the reconstructed sEMG component or leak into other contaminant components. 

\begin{figure}[t!]
   \centering
   \includegraphics[width=.75\columnwidth]{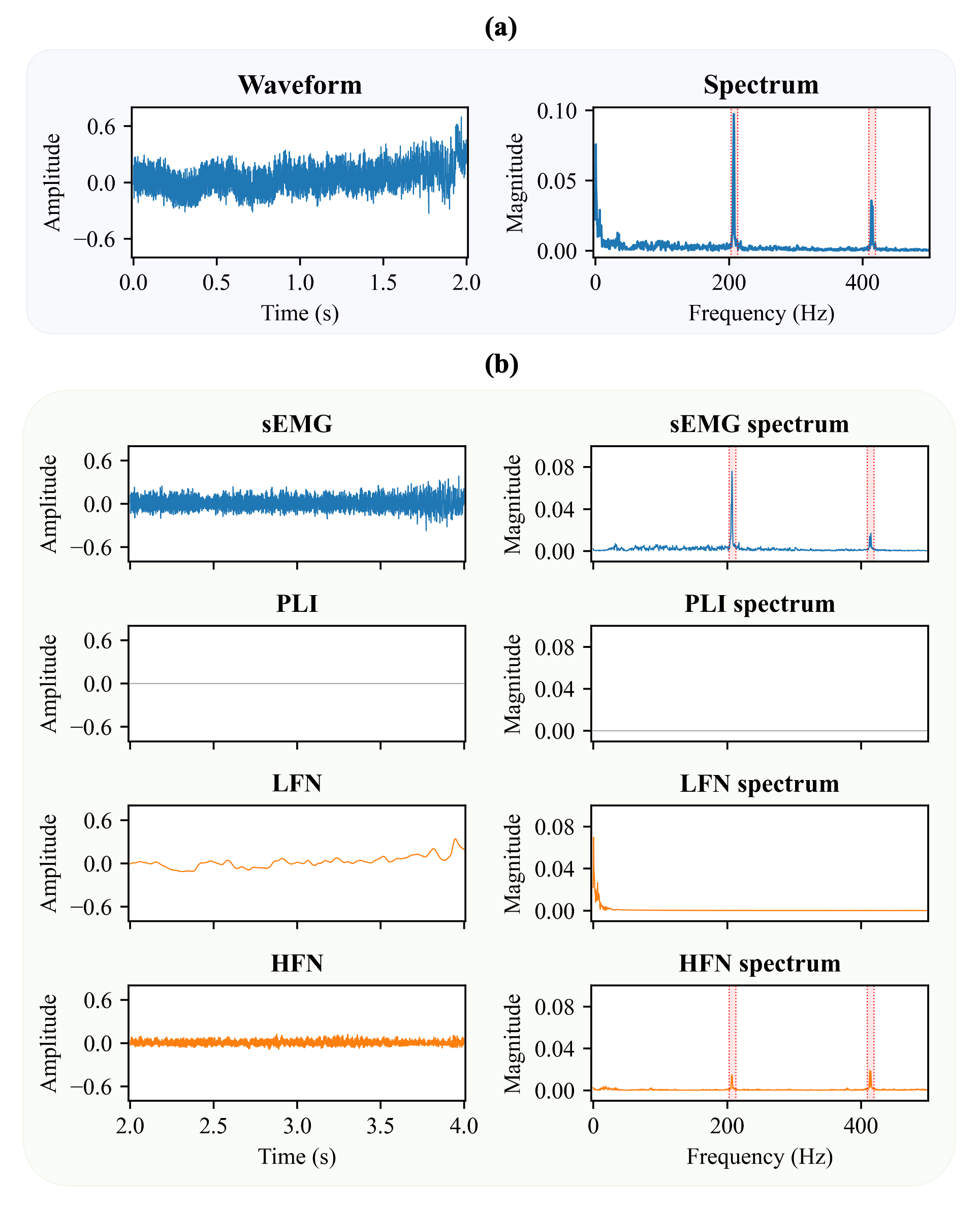}
   \caption{\textcolor{black}{Representative clinical example under out-of-distribution contamination. (a) An SQI-DB3 recording contains LFN and atypical PLI at 209.5 and 419 Hz (red shaded areas). (b) Prism-SQA assigns the low-frequency contamination primarily to LFN, whereas the atypical PLI remains partly in the reconstructed sEMG and HFN components.}}
   \label{fig:failure_case}
\end{figure}

Fig. \ref{fig:failure_case}(a) presents a representative clinical case from SQI-DB3 containing natural mixed artifacts, including LFN and atypical PLI at approximately 209.5 Hz and 419 Hz. These PLI frequencies differ substantially from the 50/60 Hz PLI and harmonics used during training. As shown in Fig. \ref{fig:failure_case}(b),  Prism-SQA assigns the low-frequency contamination primarily to the LFN component, reflecting its similarity to the LFN patterns represented during training. However, the atypical PLI is not correctly separated into the PLI component. Instead, it remains partially in the reconstructed sEMG component while leaking into the HFN component. These observations suggest that Prism-SQA can capture certain natural artifacts whose characteristics are similar to those represented during training, but may degrade when encountering out-of-distribution contaminant characteristics, leading to inaccurate quality estimation or contaminant identification.

Future work will investigate training on a broader range of real-world contaminant characteristics to improve model generalization. Another promising direction is the development of hybrid source separation frameworks that combine data-driven learning with signal-processing priors to improve robustness against unseen contaminant characteristics.

\subsubsection{Simplicity of CFV module}
The current verification module represents a proof-of-concept design that relies primarily on simple spectral or temporal fingerprints for contamination validation. While this approach effectively eliminates artifacts, its discriminative capability is restricted by the limited set of features it examines. Future work will explore applying richer contaminant-specific properties, such as periodicity for PLI and ECG or burst structure for SPK, to enhance robustness. In addition, integrating classical digital filtering techniques into the CFV may provide more nuanced refinements to the separated components than a binary pass–reject mechanism.

\begin{table}[t]
    \centering
    \scriptsize
    \caption{Computational complexity of both neural SQA methods.}
    \label{tab:computation}
    \begin{tabular}{ccccc}
     \toprule
     \multirow{2}{*}{Method} & \multirow{2}{*}{\shortstack{Parameter\\number}} & FLOPs & \multicolumn{2}{c}{Computation time (ms)}  \\
     & &  (MAC) & on GPU & on CPU \\
     \midrule
    Prism-SQA & 16.53 M & 10.6 G & 37.66 & 69.96\\
    QASE-Net & 19.05 M & 1.02 G & 6.33 & 14.82\\
    \bottomrule
    \end{tabular}
\end{table}

\subsubsection{High computational complexity for real-time or wearable usage}
Improvements in computational efficiency are needed for real-time or embedded sEMG applications. Table~\ref{tab:computation} reports the computational cost of Prism-SQA and QASE-Net for processing a 2-s sEMG segment on an NVIDIA GeForce RTX 3090 GPU and Intel Xeon Silver 4416+ CPU. While Prism-SQA remains feasible for offline analysis on high-performance hardware, it is approximately five to six times slower than QASE-Net. Potential strategies to improve efficiency include model compression through quantization or pruning \cite{han2015deep} to reduce memory footprint, as well as redesigning the architecture to support causal inference for low-latency scenarios \cite{yang2025band}. These enhancements would broaden the applicability of Prism-SQA to resource- and time-constrained environments.

\section{Conclusion}
This study presents Prism-SQA, an interpretable and adaptable neural SQA framework for sEMG. By reformulating SQA as a physiology-aware source separation and verification process, Prism-SQA addresses key limitations of black-box neural methods. The model decomposes input into an sEMG component and contaminant-specific components, which are further validated by a knowledge-driven CFV to suppress physiologically implausible artifacts. Quality indices are then computed from the verified components, decoupling metric definition from network training and enabling flexible adaptation to different clinical criteria. Experiments on three Ninapro subsets and a clinical SQI dataset demonstrate that Prism-SQA achieves competitive or superior estimation accuracy relative to a strong black-box baseline, while offering enhanced signal-level interpretability and quality-index-level adaptability. \textcolor{black}{These strengths position Prism-SQA as a transparent and customizable framework with the potential to support trustworthy sEMG SQA in clinical and research applications.}

\label{sec:conclusion}

\section*{Acknowledgment}
We thank Professor Juliano Machado from Sul-Rio-Grandense Federal Institute for providing the MOA dataset.

\section*{References}
\bibliographystyle{IEEEbib}
{\footnotesize
\bibliography{refs}}

@article{ma2020emg,
  title={{EMG signal filtering based on variational mode decomposition and sub-band thresholding}},
  author={Ma, Shihan and others},
  journal={IEEE journal of biomedical and health informatics},
  volume={25},
  number={1},
  pages={47--58},
  year={2020},
  publisher={IEEE}
}

@article{cimolato2022emg,
  title={{EMG-driven control in lower limb prostheses: A topic-based systematic review}},
  author={Cimolato, Andrea and others},
  journal={Journal of NeuroEngineering and Rehabilitation},
  volume={19},
  number={1},
  pages={43},
  year={2022},
  publisher={Springer}
}

@article{mukhopadhyay2020experimental,
  title={An experimental study on upper limb position invariant {EMG} signal classification based on deep neural network},
  author={Mukhopadhyay, Anand Kumar and Samui, Suman},
  journal={Biomedical signal processing and control},
  volume={55},
  pages={101669},
  year={2020},
  publisher={Elsevier}
}

@article{shaikh2024toward,
  title={Toward robust and accurate myoelectric controller design based on multiobjective optimization using evolutionary computation},
  author={Shaikh, Ahmed Aqeel and others},
  journal={IEEE Sensors Journal},
  volume={24},
  number={5},
  pages={6418--6429},
  year={2024},
  publisher={IEEE}
}

@article{farina2016characterization,
  title={{Characterization of human motor units from surface {EMG} decomposition}},
  author={Farina, Dario and Holobar, Ale{\v{s}}},
  journal={Proceedings of the IEEE},
  volume={104},
  number={2},
  pages={353--373},
  year={2016},
  publisher={IEEE}
}

@article{chang2012wireless,
  title={{A wireless sEMG recording system and its application to muscle fatigue detection}},
  author={Chang, Kang-Ming and Liu, Shin-Hong and Wu, Xuan-Han},
  journal={Sensors},
  volume={12},
  number={1},
  pages={489--499},
  year={2012},
  publisher={Molecular Diversity Preservation International (MDPI)}
}

@article{hogrel2005clinical,
  title={{Clinical applications of surface electromyography in neuromuscular disorders}},
  author={Hogrel, Jean-Yves},
  journal={Neurophysiologie Clinique/Clinical Neurophysiology},
  volume={35},
  number={2-3},
  pages={59--71},
  year={2005},
  publisher={Elsevier}
}

@article{sauer2024signal,
  title={Signal quality evaluation of single-channel respiratory {sEMG} recordings},
  author={Sauer, Julia and others},
  journal={Biomedical Signal Processing and Control},
  volume={87},
  pages={105414},
  year={2024},
  publisher={Elsevier}
}

@article{boyer2023reducing,
  title={Reducing Noise, Artifacts and Interference in Single-Channel {EMG} Signals: A Review},
  author={Boyer, Marianne and others},
  journal={Sensors},
  volume={23},
  number={6},
  pages={2927},
  year={2023},
  publisher={MDPI}
}

@article{farago2022review,
  title={A review of techniques for surface electromyography signal quality analysis},
  author={Farago, Emma and others},
  journal={IEEE Reviews in Biomedical Engineering},
  volume={16},
  pages={472--486},
  year={2022},
  publisher={IEEE}
}

@article{raghu2022automated,
  title={Automated biomedical signal quality assessment of electromyograms: Current challenges and future prospects},
  author={Raghu, Shriram Tallam Puranam and MacIsaac, Dawn T and Chan, Adrian DC},
  journal={IEEE Instrumentation \& Measurement Magazine},
  volume={25},
  number={1},
  pages={12--19},
  year={2022},
  publisher={IEEE}
}

@inproceedings{guo2025revisiting,
  title={Revisiting Noise Resilience Strategies in Gesture Recognition: Short-Term Enhancement in {sEMG} Analysis},
  author={Guo, Weiyu and others},
  booktitle={Proc. ICML},
  year={2025}
}

@article{zhao2024biosignal,
  title={A biosignal quality assessment framework for high-density {sEMG} decomposition},
  author={Zhao, Zeming and others},
  journal={Biomedical Signal Processing and Control},
  volume={90},
  pages={105800},
  year={2024},
  publisher={Elsevier}
}

@inproceedings{fraser2011cleanemg,
  title={{CleanEMG—Power line interference estimation in sEMG using an adaptive least squares algorithm}},
  author={Fraser, Graham D and others},
  booktitle={Proc. EMBC},
  year={2011},
}

@article{abser2011cleanemg,
  title={{CleanEMG: Quantifying power line interference in surface {EMG} signals}},
  author={Abser, Nurul and others},
  journal={Proc. CMBES},
  year={2011}
}

@article{abser2012cleanemg,
  title={{CleanEMG: Comparing interpolation strategies for power line interference quantification in surface {EMG} signals}},
  author={Abser, Nurul and others},
  journal={Proc. CMBES},
  year={2012}
}

@article{sinderby1995automatic,
  title={{Automatic assessment of electromyogram quality}},
  author={Sinderby, Christer and Lindstrom, Lars and Grassino, AE},
  journal={Journal of Applied Physiology},
  volume={79},
  number={5},
  pages={1803--1815},
  year={1995}
}

@article{keshtkaran2014fast,
  title={{A fast, robust algorithm for power line interference cancellation in neural recording}},
  author={Keshtkaran, Mohammad Reza and Yang, Zhi},
  journal={Journal of neural engineering},
  volume={11},
  number={2},
  pages={026017},
  year={2014},
  publisher={IOP Publishing}
}

@inproceedings{chang2020assessment,
  title={{Assessment of {EMG} benchmark data for gesture recognition using the NinaPro database}},
  author={Chang, Jason and Phinyomark, Angkoon and Scheme, Erik},
  booktitle={Proc. EMBC},
  year={2020},
}

@article{oo2020signal,
  title={{Signal-to-noise ratio estimation in electromyography signals contaminated with electrocardiography signals}},
  author={Oo, Thandar and Phukpattaranont, Pornchai},
  journal={Fluctuation and Noise Letters},
  volume={19},
  number={03},
  pages={2050027},
  year={2020},
  publisher={World Scientific}
}

@inproceedings{lee2024non,
  title={{A Non-Intrusive Neural Quality Assessment Model for Surface Electromyography Signals}},
  author={Lee, Cho-Yuan and others},
  booktitle={Proc. EMBC},
  year={2024},
}

@inproceedings{del2011assessment,
  title={Assessment of different methods to estimate electrocardiogram signal quality},
  author={Del Rio and others},
  booktitle={Proc. CinC},
  year={2011},
}

@article{machado2021deep,
  title={{Deep learning for surface electromyography artifact contamination type detection}},
  author={Machado, Juliano and Machado, Amauri and Balbinot, Alexandre},
  journal={Biomedical Signal Processing and Control},
  volume={68},
  pages={102752},
  year={2021},
  publisher={Elsevier}
}

@article{salahuddin2022transparency,
  title={{Transparency of deep neural networks for medical image analysis: A review of interpretability methods}},
  author={Salahuddin, Zohaib and others},
  journal={Computers in biology and medicine},
  volume={140},
  pages={105111},
  year={2022},
  publisher={Elsevier}
}

@article{quinn2022three,
  title={{The three ghosts of medical AI: Can the black-box present deliver?}},
  author={Quinn, Thomas P and others},
  journal={Artificial intelligence in medicine},
  volume={124},
  pages={102158},
  year={2022},
  publisher={Elsevier}
}

@article{cuadros2022automatic,
  title={Automatic detection of poor quality signals as a pre-processing scheme in the analysis of {sEMG} in swallowing},
  author={Cuadros-Acosta, J and Orozco-Duque, Andr{\'e}s},
  journal={Biomedical Signal Processing and Control},
  volume={71},
  pages={103122},
  year={2022},
  publisher={Elsevier}
}

@article{hochreiter1997long,
  title={{Long short-term memory}},
  author={Hochreiter, Sepp and Schmidhuber, J{\"u}rgen},
  journal={Neural computation},
  volume={9},
  number={8},
  pages={1735--1780},
  year={1997},
  publisher={MIT press}
}

@inproceedings{wang2023ecg,
  title={{ECG} Artifact Removal from Single-Channel Surface {EMG} Using Fully Convolutional Networks},
  author={Wang, Kuan-Chen and others},
  booktitle={Proc. ICASSP},
  year={2023}
}

@inproceedings{kingma2015adam,
  title={Adam: A method for stochastic optimization},
  author={Kingma, Diederik P and Ba, Jimmy},
  booktitle={Proc. ICLR},
  year={2015}
}

@article{atzori2014electromyography,
  title={Electromyography data for non-invasive naturally-controlled robotic hand prostheses},
  author={Atzori, Manfredo and others},
  journal={Scientific data},
  volume={1},
  number={1},
  pages={1--13},
  year={2014},
  publisher={Nature Publishing Group}
}

@inproceedings{zhang2023semg,
  title={{sEMG} Signal Denoising Based on Variational Mode Decomposition and Wavelet Thresholding},
  author={Zhang, Chen and Zhou, Zijian and Zhou, Shuren},
  booktitle={Proc. ICMSP},
  year={2023}
}

@article{yadav2023noise,
  title={Noise confiscation from {sEMG} through enhanced adaptive filtering based on evolutionary computing},
  author={Yadav, Shubham and others},
  journal={Circuits, Systems, and Signal Processing},
  pages={1--33},
  year={2023},
  publisher={Springer}
}

@inproceedings{mateo2008neural,
  title={{Neural network based canceller for powerline interference in ECG signals}},
  author={Mateo, J and others},
  booktitle={Proc. CinC},
  year={2008}
}

@article{goldberger2000physiobank,
  title={{PhysioBank, PhysioToolkit, and PhysioNet: components of a new research resource for complex physiologic signals}},
  author={Goldberger, Ary L and others},
  journal={circulation},
  volume={101},
  number={23},
  pages={e215--e220},
  year={2000},
  publisher={Am Heart Assoc}
}

@article{mccool2014identification,
  title={Identification of contaminant type in surface electromyography (EMG) signals},
  author={McCool, Paul and others},
  journal={IEEE Transactions on Neural Systems and Rehabilitation Engineering},
  volume={22},
  number={4},
  pages={774--783},
  year={2014},
  publisher={IEEE}
}

@inproceedings{abdelazez2018detection,
  title={Detection of noise type in electrocardiogram},
  author={Abdelazez, Mohamed and Rajan, Sreeraman and Chan, Adrian DC},
  booktitle={Proc. MeMeA},
  year={2018}
}

@article{moody1984noise,
  title={A noise stress test for arrhythmia detectors},
  author={Moody, George B and Muldrow, W and Mark, Roger G}, 
  journal={Computers in cardiology},
  volume={11},
  number={3},
  pages={381--384},
  year={1984}
}

@inproceedings{li2017wavelet,
  title={{Wavelet-based detection on MUAPs decomposed from sEMG under different levels of muscle isometric contraction}},
  author={Li, Ziyou and others},
  booktitle={Proc. ROBIO},
  year={2017},
}

@article{cao2022control,
  title={Control of external devices based on {sEMG} signals},
  author={Cao, Fangqi and others},
  journal={Academic Journal of Computing \& Information Science},
  volume={5},
  number={14},
  pages={126--132},
  year={2022},
  publisher={Francis Academic Press}
}

@article{thongpanja2013mean,
  title={Mean and median frequency of {EMG} signal to determine muscle force based on time-dependent power spectrum},
  author={Thongpanja, Sirinee and others},
  journal={Elektronika ir Elektrotechnika},
  volume={19},
  number={3},
  pages={51--56},
  year={2013}
}

@inproceedings{xiaojing2011feature,
  title={Feature extraction and classification of {sEMG} based on {ICA} and {EMD} decomposition of {AR} model},
  author={Xiaojing, Shang and Yantao, Tian and Yang, Li},
  booktitle={Proc. ICECC},
  year={2011}
}

@inproceedings{han2015deep,
  title={Deep Compression: Compressing Deep Neural Network with Pruning, Trained Quantization and Huffman Coding},
  author={Han, Song and Mao, Huizi and Dally, William J},
  booktitle={Proc. ICLR},
  year={2016}
}

@inproceedings{yang2025band,
  title={{Band-SCNet}: A Causal, Lightweight Model for High-Performance Real-Time Music Source Separation},
  author={Yang, Junqi and others},
  booktitle={Proc. Interspeech},
  year={2025}
}

@article{huang1998empirical,   title={The empirical mode decomposition and the Hilbert spectrum for nonlinear and non-stationary time series analysis},   author={Huang, Norden E and others},   journal={Proceedings of the Royal Society of London. Series A: mathematical, physical and engineering sciences},   volume={454},   number={1971},   pages={903--995},   year={1998},   publisher={The Royal Society} }

@inproceedings{venkatesh2024real,
  title={Real-time low-latency music source separation using hybrid spectrogram-tasnet},
  author={Venkatesh, Satvik and others},
  booktitle={Proc. ICASSP},
  year={2024},
}

@article{alizadegan2025comparative,
  title={{Comparative study of long short-term memory (LSTM), bidirectional LSTM, and traditional machine learning approaches for energy consumption prediction}},
  author={Alizadegan, Hamed and others},
  journal={Energy Exploration \& Exploitation},
  volume={43},
  number={1},
  pages={281--301},
  year={2025},
  publisher={SAGE Publications Sage UK: London, England}
}

@article{defossez2019music,
  title={Music source separation in the waveform domain},
  author={D{\'e}fossez, Alexandre and others},
  journal={arXiv preprint arXiv:1911.13254},
  year={2019}
}

@inproceedings{le2019sdr,
  title={{SDR--half-baked or well done?}},
  author={Le Roux, Jonathan and others},
  booktitle={Proc. ICASSP},
  year={2019},
}

@inproceedings{guso2022loss,
  title={On loss functions and evaluation metrics for music source separation},
  author={Gus{\'o}, Enric and others},
  booktitle={Proc. ICASSP},
  year={2022},
}

@inproceedings{kong2018joint,
  title={A joint separation-classification model for sound event detection of weakly labelled data},
  author={Kong, Qiuqiang and others},
  booktitle={Proc. ICASSP},
  year={2018},
}

@article{karamatli2019audio,
  title={Audio source separation using variational autoencoders and weak class supervision},
  author={Karamatl{\i}, Ertu{\u{g}} and Cemgil, Ali Taylan and K{\i}rb{\i}z, Serap},
  journal={IEEE Signal Processing Letters},
  volume={26},
  number={9},
  pages={1349--1353},
  year={2019},
  publisher={IEEE}
}

@inproceedings{ronneberger2015u,
  title={{U-net: Convolutional networks for biomedical image segmentation}},
  author={Ronneberger, Olaf and Fischer, Philipp and Brox, Thomas},
  booktitle={Proc. MICCAI},
  year={2015},
}

@article{wang2024trustemg,
  title={{TrustEMG-Net: Using representation-masking transformer with U-net for surface electromyography enhancement}},
  author={Wang, Kuan-Chen and others},
  journal={IEEE Journal of Biomedical and Health Informatics},
  volume={29},
  number={4},
  pages={2506--2520},
  year={2024},
  publisher={IEEE}
}

@inproceedings{takahashi2018mmdenselstm,
  title={{MMDenseLSTM: An efficient combination of convolutional and recurrent neural networks for audio source separation}},
  author={Takahashi, Naoya and Goswami, Nabarun and Mitsufuji, Yuki},
  booktitle={Proc. IWAENC},
  year={2018},
}

@article{spearman1961general,
  title={{'General Intelligence' Objectively Determined and Measured.}},
  author={Spearman, Charles},
  year={1961},
  publisher={Appleton-Century-Crofts}
}

\end{document}